\documentclass[12pt]{article}

\usepackage{scicite}
\usepackage{graphicx}
\usepackage{times}

\usepackage{color}
\usepackage{ulem}
\usepackage[dvipdfmx]{hyperref}

\newenvironment{sciabstract}{%
\begin{quote} \bf}
{\end{quote}}

\newcommand{\rev}{\textcolor{black}}

\title{Mapping Solar Magnetic Fields from \\the Photosphere to the Base of the Corona}

\author
{Ryohko Ishikawa$^{1\ast}$, 
 Javier Trujillo Bueno$^{2,3,4}$,\\
 Tanaus\'u del Pino Alem\'an$^{2,3}$,
 Takenori J. Okamoto$^{1}$,
 David E. McKenzie$^{5}$,\\
 Fr\'ed\'eric Auch\`ere$^{6}$,
 Ryouhei Kano$^{1}$,
 Donguk Song$^{1}$,
 Masaki Yoshida$^{1,7}$,\\
 Laurel A. Rachmeler$^{8}$,
 Ken Kobayashi$^{5}$,
 Hirohisa Hara$^{1}$,
 Masahito Kubo$^{1}$,\\
 Noriyuki Narukage$^{1}$,
Taro Sakao$^{9}$,
Toshifumi Shimizu$^{9}$,
Yoshinori Suematsu$^{1}$, \\
Christian Bethge$^{\rev{10}}$,
 Bart De Pontieu$^{11,12,13}$,
Alberto Sainz Dalda$^{11,14,15}$,\\
Genevieve D. Vigil$^{5,\rev{16}}$,
 Amy Winebarger$^{5}$,\\
 Ernest Alsina Ballester$^{\rev{17}}$,
  Luca Belluzzi$^{\rev{17,18}}$,
 Ji\v{r}\'i  \v{S}t\v{e}p\'an$^{\rev{19}}$,\\
 Andr\'es Asensio Ramos$^{2,3}$,
 Mats Carlsson$^{12,13}$,
 Jorrit Leenaarts$^{\rev{20}}$
\\
\normalsize{$^{1}$National Astronomical Observatory of Japan, Mitaka, Tokyo 181-8588 Japan}\\
\normalsize{$^{2}$Instituto de Astrof\'\i sica de Canarias, 38205 La Laguna, Tenerife, Spain}\\
\normalsize{$^{3}$Departamento de Astrof\'isica, Universidad de La Laguna, 
E-38206 La Laguna, Tenerife, Spain}\\
\normalsize{$^{4}$Consejo Superior de Investigaciones Cient\'ificas, Spain}\\
\normalsize{$^{5}$NASA Marshall Space Flight Center, Huntsville, AL 35812, USA}\\
\normalsize{$^{6}$Institut d'Astrophysique Spatiale, 91405 Orsay Cedex, France}\\
\normalsize{$^{7}$Department of Astronomical Science, School of Physical Sciences,}\\
\normalsize{SOKENDAI (The Graduate University for Advanced Studies)}\\
\normalsize{$^{8}$National Oceanic and Atmospheric Administration,}\\ 
\normalsize{National Centers for Environmental Information, Boulder, CO 80305, USA}\\
\normalsize{$^{9}$Institute of Space and Astronautical Science, 
Japan Aerospace Exploration Agency,}\\
\normalsize{Sagamihara, Kanagawa 252-5210, Japan}\\
\normalsize{\rev{$^{10}$Cooperative Institute for Research in Environmental Sciences,}}\\ 
\normalsize{\rev{University of Colorado at Boulder, Boulder, CO 80305, USA}}\\
\normalsize{$^{11}$Lockheed Martin Solar \& Astrophysics Laboratory, Palo Alto, CA 94304, USA}\\
\normalsize{$^{12}$Rosseland Centre for Solar Physics, University of Oslo, NO-0315 Oslo, Norway}\\ 
\normalsize{$^{13}$Institute of Theoretical Astrophysics, University of Oslo, NO-0315 Oslo, Norway}\\
\normalsize{$^{14}$Bay Area Environmental Research Institute, Moffett Field, CA 94035, USA}\\
\normalsize{$^{15}$Stanford University, HEPL, Stanford, CA 94305-4085, USA}\\
\normalsize{\rev{$^{16}$}Universities Space Research Association, Huntsville, AL 35805, USA}\\
\normalsize{\rev{$^{17}$}Istituto Ricerche Solari Locarno, CH - 6605 Locarno Monti, Switzerland}\\
\normalsize{\rev{$^{18}$}Leibniz-Institut f\"ur Sonnenphysik (KIS), Sch\"oneckstr. 6, D-79104, Freiburg, Germany}\\
\normalsize{\rev{$^{19}$}Astronomical Institute, Academy of Sciences of the Czech Republic,}\\ 
\normalsize{25165 Ondrejov, Czech Republic}\\
\normalsize{\rev{$^{20}$}Institute for Solar Physics, Department of Astronomy, Stockholm University,}\\ 
\normalsize{AlbaNova University Centre, SE-106 91, Stockholm, Sweden}\\
\normalsize{$^\ast$To whom correspondence should be addressed; E-mail:  ryoko.ishikawa@nao.ac.jp}
}

\date{}

\begin{document} 


\baselineskip24pt


\maketitle


\begin{sciabstract}

Routine ultraviolet imaging of the Sun's upper atmosphere shows the 
spectacular manifestation of solar activity; yet we remain blind to its main driver, the magnetic field. 
Here we report unprecedented spectropolarimetric observations 
of an active region plage and its surrounding enhanced network, 
showing circular polarization in ultraviolet (Mg {\sc ii} $h$ \& $k$ and Mn {\sc i})  
and visible (Fe {\sc i}) lines. We infer the longitudinal magnetic field from the photosphere to the very 
upper chromosphere. At the top of the plage chromosphere the field strengths reach more than 300
gauss, strongly correlated with the Mg {\sc ii} $k$ line core intensity 
and the electron pressure. This unique mapping  
shows how the magnetic field couples the different atmospheric layers and reveals 
the magnetic origin of the heating in the plage chromosphere.

\end{sciabstract}

\section*{\rev{125-character teaser}}
\rev{A novel space experiment achieves an unprecedented map of solar magnetic fields 
from the photosphere to the base of the corona}

\section*{Main Text}
\subsection*{\rev{Introduction}}
The chromosphere is a very important region of the solar atmosphere, with an extension 
of several thousand km, located between the relatively 
cool surface layers of the photosphere and the overlying 
hot corona \cite{2009ASPC..405..157H,2012RSPTA.370.3129R,2019ARA&A..57..189C}.
Although the temperature of the chromospheric plasma does not exceed $10^4$ K, 
the fact that its density is much larger 
than that of the extended and rarified corona implies 
that much more mechanical energy is required 
to sustain the chromosphere than the million-degree corona. 
Moreover, from the visible photospheric  
surface to the chromosphere-corona transition region (TR) 
the plasma density decreases exponentially 
by several orders of magnitude, more rapidly than the magnetic field strength. As a result,  
the $\beta=1$ corrugated surface, where the ratio of gas to magnetic pressure is unity,  
lies inside the chromosphere. Above the $\beta=1$ surface 
the magnetic field essentially dominates  
the structuring and dynamics of the plasma. 
This, together with the fact that the non-thermal energy 
needed to heat the corona must propagate through the chromosphere, 
explains why it is indeed a crucial interface region to
solve many of the key problems in solar and stellar physics.

It is impossible to fully understand the solar chromosphere 
without mapping its magnetic structure, especially in  
the relatively hot layers of the upper chromosphere 
and TR where $\beta<1$ 
\cite{2009ASPC..405..157H,2012RSPTA.370.3129R,2019ARA&A..57..189C,
2007cemf.book.....P,2005ApJ...618.1020G,2007Sci...318.1574D,
2008ApJ...679L..57I,2013Natur.493..501C,2017Sci...356.1269M}.
To this end, we need to measure and model the polarization of ultraviolet (UV) spectral lines 
originating in such atmospheric regions
\cite{2017SSRv..210..183T}. The theoretical investigations reported 
in the just quoted review paper led us to a series of suborbital 
space experiments called CLASP, which required the 
development of novel instrumentation (supplementary material 1A). 
The present investigation is based on a unique dataset acquired by   
the Chromospheric LAyer Spectropolarimeter (CLASP2), a suborbital space experiment that 
on 11 April 2019 allowed us to measure the first ever spectrally resolved Stokes profiles 
across the Mg {\sc ii} $h$ \& $k$ lines in active and quiet regions of the solar disk.

\subsection*{\rev{Methods}}

We focus on the CLASP2 measurements in 
an active region plage and its surrounding enhanced 
network (see Figures 1A and 1B, and note the position of the spectrograph's slit). 
These are poorly understood regions of the solar disk with 
large concentrations of magnetic flux, located at the footpoints of 
many coronal loops through which the mechanical energy that energizes the corona 
propagates \cite{2019ARA&A..57..189C,2010MmSAI..81..604K}. 
Although CLASP2 successfully measured 
the four Stokes parameters of several lines around 280 nm 
(see Figure S2), in this paper we use only the intensity (Stokes $I$) 
and circular polarization (Stokes $V$)  
profiles of the Mg {\sc ii} $h$ \& $k$ lines and of two nearby Mn {\sc i} lines. 
The observed $V/I$ spectra of these four lines can be seen in panel D of Figure 1, 
which also shows the detection of weaker circular polarization signals in other spectral lines.
\rev{We point out that in this paper we always deal with the fractional circular polarization 
$V(\lambda)/I(\lambda)$ (hereafter, $V/I$), with $\lambda$ the wavelength.} 

Interestingly, CLASP2 detected clear circular polarization 
signals not only within the bright region of the plage,
but also at the enhanced network elements located at {\bf d} and {\bf e} (Figure 1A).
The Stokes $V/I$ profiles of the Mg {\sc ii} $h$ \& $k$ 
resonance lines have two external 
lobes and two inner lobes, which encode information on the longitudinal component 
of the magnetic field in 
the middle chromosphere and at the top of the upper chromosphere, respectively 
(supplementary materials 2A$-$2C).
The $V/I$ profiles of the two Mn {\sc i} lines at 279.91~nm 
and at 280.19~nm 
have only two lobes, which provide information 
on the longitudinal field component in the lower chromosphere (supplementary material 
2D). In addition, in order to obtain the longitudinal field component in the 
underlying photosphere, we have used the Stokes profiles of two visible Fe {\sc i} lines measured 
by the Hinode spacecraft during the CLASP2 flight (supplementary material 1B). 

All such circular polarization signals are produced 
by the Zeeman effect caused by the magnetic field 
that permeates the solar atmosphere. 
To determine the longitudinal field component from 
the Stokes $I$ and $V$ profiles measured by CLASP2 
in the above-mentioned chromospheric lines 
we applied the weak field approximation 
(WFA; supplementary material 1D). Figure 2 shows an example of 
the observed profiles, with the colored curves indicating the corresponding WFA fits.  
The generally stronger 
photospheric magnetic field values reported here result from 
the application of a suitable inversion code 
to the Stokes profiles observed by the Hinode spacecraft (supplementary material 1E).

\subsection*{\rev{Results and Discussion}}
Figure 3 gives the variation, along the spatial direction of the spectrograph's slit, of the 
longitudinal component of the magnetic field. 
The figure shows this quantity in the photosphere (green curves), in the 
lower chromosphere (blue symbols), in the middle 
chromosphere (black symbols) and at the top of the upper chromosphere 
(red symbols). Note that the error bars ($1\sigma$) of the 
inferred values, due to the noise in $V/I$, 
are very small, thanks to the high-precision CLASP2 measurements.
The observed bright plage region extends from $-98$~arcsec to about 28~arcsec, 
with a small dark region located around $-62$~arcsec (see Figure 1A). 

In the bright plage region, we see clearly in Figure 3 
that the magnetic field 
always shows a single magnetic polarity, except perhaps at the    
$-65$~arcsec and $+12$~arcsec positions,   
where the retrieved opposite polarity fields may not be statistically significant 
because they are comparable to or smaller than the $1\sigma$ uncertainty.
As expected, the strongest magnetic fields are found in the photosphere of 
the observed plage, where the magnetic field appears to be organized into small regions with 
strong magnetic concentrations (with longitudinal field components as large as 1250~G) 
separated by small regions with longitudinal field values of the order of 10~G
(see the green curve and its spatial fluctuation, which has typical 
scales of less than 10~arcsec). In the lower chromosphere 
(blue symbols) the magnetic field also 
shows a substantial fluctuation on similar spatial scales, 
but the amplitude of the spatial variation is significantly smaller and this variation 
is not exactly in phase with the one in the 
underlying photosphere 
(e.g., note that the spatial locations of the blue-symbol peaks 
do not coincide exactly with those of the green curve). In the low 
chromosphere the maximum longitudinal 
magnetic field values are about 700~G (blue symbols).
In the middle chromosphere of the plage and at the top of its upper chromosphere   
the black and red symbols of Figure 3 indicate that 
the longitudinal component of the magnetic field varies from 
almost zero gauss (slit positions around $+12$ and $-65$~arcsec) to  
more than 300~G (locations {\bf a}, {\bf c} and $-44$~arcsec),  
showing a smoother  
spatial variation than in the lower chromosphere (blue symbols) 
and photosphere (green curve).  
The longitudinal magnetic field values found near 
the TR (red symbols, inferred from the inner lobes of the Stokes $V/I$ profiles of Mg {\sc ii}  
$h$ \& $k$) are not much weaker than those found in the middle chromosphere 
(black symbols, determined from the outer $V/I$ lobes of the $h$ line), although it must be noted  
that the latter values are a lower limit (supplementary 
material 2C). Finally, note that at almost all of the spatial positions 
where the fluctuating green curve reaches its local minima, the photospheric 
longitudinal field values are smaller than at the same locations in the overlying layers    
(e.g., see position {\bf b} and the corresponding Figure S5).
However, at the spatial positions where the longitudinal component of the  
magnetic field in the lower chromosphere (blue symbols) reaches its smallest values,  
we find similar ones in the middle chromosphere and at the top of the upper chromosphere
(black and red symbols, respectively). These and the previously mentioned fact, 
namely, that the spatial fluctuations of the longitudinal field 
values decrease with height, confirm that the plage magnetic field is highly structured 
in the photosphere, with strong and weak field variations, 
and that \rev{it expands rapidly merging and spreading horizontally  
in the overlying chromosphere} where 
the field is weaker and has smoother spatial variations.

We have found a remarkably high correlation between the longitudinal component of the magnetic field 
in the middle and top layers of the plage chromosphere and the intensity 
at the center and the emission peaks of the Mg {\sc ii} $k$ line (see Figure 3).
As shown in Table 1, the coefficient that quantifies this correlation increases with height in the plage 
atmosphere, and it is important to note that 
the intensity in the $k$ line-core emission peaks correlates with the temperature 
at its formation height \cite{2013ApJ...772...90L}.
Moreover, from inversions of the observed Mg {\sc ii} $h$ \& $k$ intensity 
profiles (supplementary material 2E), we have found (see Table 1) that in the middle and the 
upper chromosphere the longitudinal component of the magnetic field is also significantly correlated 
with the electron pressure (i.e., with the product of the temperature and the electron density).
These pieces of empirical information strongly support that the heating 
of the upper chromosphere of active region plages is of magnetic origin.

Consider now the supergranulation cell outlined by its bright 
chromospheric network, which is located above 60~arcsec in Figure~1A.
The slit of the spectrograph crosses network features  
at 63~arcsec (location {\bf d}) and at 97~arcsec (location {\bf e}).
As shown by the green curve 
of Figure 3, the longitudinal component of the 
magnetic field in the photosphere of these two network 
elements is about 220~G. At these network locations 
we find longitudinal field strengths of about    
160~G and 80~G in the lower and middle chromosphere, respectively 
(see Figure 4 for location {\bf d}). Figure 3 shows that the 
magnetic fields in the atmospheres of these two 
network elements have opposite polarities.
As seen in the top panels of Figure 4, the inner lobes of the Stokes $V/I$ 
profiles of the $h$ \& $k$ lines are so weak that their circular polarization 
amplitudes are compatible with zero within the error bars, 
and this is not due to cancelations of the 
circular polarization signals during the 150.4 s integration time 
(supplementary material 1D). The absence 
of significant inner lobes in such $V/I$ profiles
leads us to conclude that the magnetic field in the very upper chromosphere 
of the observed network features is weaker than about 10~G.
Therefore, in the atmospheres of these network elements 
the longitudinal field values vary approximately 
from 220~G in the photosphere, to 160~G in the lower chromosphere, to at least 
80~G in the middle chromosphere,  
and to less than 10~G at the top of the upper chromosphere.

We foresee three possible scenarios compatible with the above-mentioned results from the
CLASP2 spectropolarimetric observations of the network features:   
(a) the magnetic loops of the network do not reach 
the very top of the chromosphere,
and return to the photosphere after reaching 
chromospheric heights, (b) the magnetic fields in the very upper chromosphere 
of the network regions are too weak 
to be detected through the Zeeman effect in the Mg {\sc ii} 
resonance lines, and (c) the magnetic field in the top layers of the network chromosphere
becomes nearly perpendicular to the line of sights of the 
CLASP2 observation, so that there is hardly any longitudinal field component to measure. 
Our \rev{CLASP2 circular polarization measurements} provide quantitative evidence that the commonly held 
picture of a wineglass-shaped magnetic canopy of network fields that fill the entire quiet 
solar chromosphere above a certain height is too simplistic to describe the expansion     
of the magnetic field in the observed enhanced network regions 
\cite{2003ApJ...597L.165S,2009odsm.book..317W}. 
\rev{An advanced inversion code is currently under development, and we hope that its 
future application to the four Stokes profiles observed by CLASP2 (see Figure S2) will allow us to 
disentangle among the three options mentioned above.}

Although a suborbital space experiment can only provide a few minutes of observing time, the 
unprecedented observations achieved by CLASP2 have 
demonstrated the enormous diagnostic potential of spectropolarimetry 
in the spectral region of the Mg {\sc ii} $h$ \& $k$ lines. 
Of particular interest are the circular polarization signals 
observed in the Mg {\sc ii} and Mn {\sc i}  
resonance lines. This is the first time that the 
longitudinal component of the magnetic field is simultaneously determined 
at several heights from the photosphere to the very top layers of the 
chromosphere of active region plages 
and enhanced network features, at each position along the spatial direction of the 
spectrograph's slit.
This opens up the possibility of mapping the magnetic field throughout the solar 
atmosphere over large fields of view, an empirical information that is crucial for 
obtaining more accurate coronal field estimates and for  
deciphering how the magnetic field couples the different atmospheric layers and 
channels the mechanical energy that heats 
the chromosphere and the corona of our nearest star.


\section*{Acknowledgments}
CLASP2 is an international partnership between NASA/MSFC, NAOJ, 
JAXA, IAC, and IAS; additional partners include ASCR, IRSOL, LMSAL, 
and the University of Oslo.
The Japanese participation was funded by \rev{JAXA} 
as a Small Mission-of-Opportunity Program, JSPS KAKENHI 
Grant numbers JP25220703 and JP16H03963, 2015 ISAS Grant 
for Promoting International Mission Collaboration, and by 2016 NAOJ 
Grant for Development Collaboration. The USA participation 
was funded by NASA Award 16-HTIDS16\_2-0027. 
The Spanish participation was funded by the European 
Research Council (ERC) under the European Union's Horizon 2020 
research and innovation programme (Advanced Grant agreement No. 742265). 
The French hardware participation was funded by CNES funds CLASP2-13616A and 13617A.
The CLASP2 team acknowledges Dr. Shin-nosuke Ishikawa, who led the development 
of the critical component of the polarization modulation unit (PMU) at ISAS/JAXA.
Hinode is a Japanese mission developed and launched by ISAS/JAXA, 
with NAOJ as domestic partner and NASA and STFC (UK) as international partners. 
It is operated by these agencies in cooperation with ESA and NSC (Norway).
IRIS is a NASA Small Explorer Mission developed and operated by LMSAL 
with mission operations executed at NASA Ames Research Center 
and major contributions to downlink communications funded by ESA and the Norwegian Space Centre.

\section*{Funding}
R.I. acknowledges the NAOJ Overseas Visit Program for Young Researchers (FY2019) 
and a Grant-in-Aid for Early-Career Scientists JP19K14771. J.T.B. acknowledges 
the funding received from the European Research Council  
through Advanced Grant agreement No. 742265.  
E.A.B. and L.B. acknowledge the funding received from the Swiss National Science Foundation   
through grants 200021-175997 and CRSII5-180238. B.D.P. and A.S.D. 
acknowledge support from NASA contract NNG09FA40C (IRIS). 
\rev{J. \v{S}. acknowledges financial support by the Grant Agency of the Czech
Republic through grant 19-20632S and project RVO:67985815.}
M.C. acknowledges support 
from the Research Council of Norway through its Centres of Excellence scheme, 
project number 262622, and through grants of computing time from the Programme for Supercomputing.

\section*{Author contributions}
R.I. and J.T.B conceived the investigation, critically discussed each development step 
and wrote the manuscript. R.I. led the data analysis, while    
J.T.B and T.dP.A. led the theoretical investigations applied to this study, 
which included discussions with E.A.B., L.B., J.\v{S}. and A.A.R. The comparison 
between the CLASP2 and IRIS data was done by T.J.O., who also 
processed the Hinode/SOT data. D.E.M., R.I, J.T.B, and F.A.  
are the CLASP2 Principal Investigators from USA, Japan, Spain, and France, respectively.
T.J.O and L.A.R., the CLASP2 project scientists, helped to   
coordinate the observations with IRIS and Hinode. B.D.P. conducted the IRIS observation
and together with A.S.D. performed the inversions based on the IRIS$^2$ database.
C.B., T.J.O. and D.S. contributed to the data calibration. R.K., D.S., M.Y., K.K.,   
H.H., M.K., N.N., T.S., T.S., Y.S., G.D.V., and A.W. 
contributed to the development of the instrument.
M.C. and J.L. provided information on models of the solar chromosphere.

\section*{Competing interests}
There are no competing interests.

\section*{Data and materials availability}
\rev{All data needed to evaluate our conclusions
are presented in the paper and/or in the Supplementary Materials.}
The calibrated CLASP2 data will be made publicly available soon through the Virtual Solar Observatory.
The Level 2 Hinode/SOT-SP data are available at https://csac.hao.ucar.
edu/sp\_data.php,  
the IRIS data at https://iris.lmsal.com/data.html, and the SDO data at http://jsoc.
stanford.edu.


\clearpage
\noindent {\bf \rev{Fig. 1}. CLASP2 and Hinode data.} 
{\bf (A)} Broadband Lyman-$\alpha$ image obtained by the CLASP2 slit-jaw system.
The red line indicates the radially oriented 
slit of the CLASP2 spectrograph, which covers 196 arcsec. 
{\bf (B)} Longitudinal component of the photospheric magnetic field inferred from the 
Stokes profiles observed by Hinode/SOT-SP in Fe {\sc i} visible lines.
{\bf (C)} Variation along the slit of the 
\rev{intensity profile $I(\lambda)$ observed in the spectral region of the Mg {\sc ii} $h$ \& $k$ lines.}
\rev{The labels at the top of panel C indicate the 
location of Mg {\sc ii} $k$ at 279.64~nm, Mg {\sc ii} $h$ 
at 280.35~nm, and the Mn {\sc i} lines at 279.91~nm and 280.19~nm.}
{\bf (D)} Fractional 
circular polarization \rev{$V(\lambda)/I(\lambda)$} observed by CLASP2 around 280~nm. 
The $I$ and $V/I$ spectra are the result of temporally averaging 
the individual Stokes parameters during 150.4~s. 
\rev{The vertical axes indicate the distance in arcseconds along the spatial 
direction of the CLASP2 slit, measured from its center.} 
The small gap seen in panels C and D results from the lack of data in a few deteriorated pixels. 
\begin{figure}[t]
\begin{center}
\includegraphics[keepaspectratio,width=140mm]{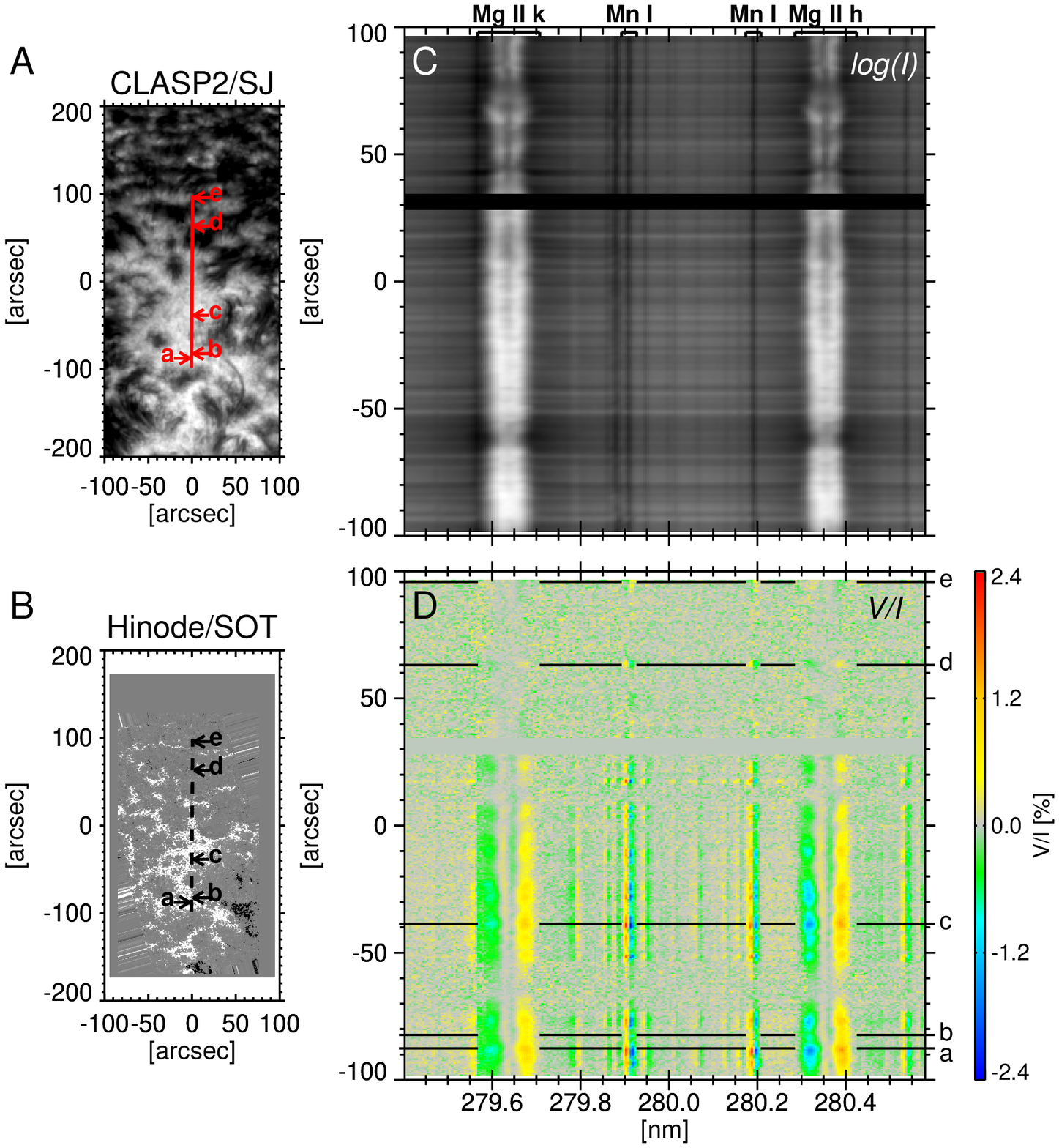}
\end{center}
\end{figure}

\clearpage

\clearpage
\noindent {\bf \rev{Fig. 2.} Example of the Stokes $I$ and $V/I$ profiles observed by 
CLASP2.} The profiles correspond to
Mg {\sc ii} $k$ at 279.64~nm, Mg {\sc ii} $h$ at 280.35~nm, 
Mn {\sc i} at 279.91~nm, and Mn~{\sc i} at 280.19~nm
at location {\bf c} in Fig.~1, where the longitudinal field retrieved  
from the Mg~{\sc ii} $h$ \& $k$ lines is among the strongest ones.
The gray curves show the
corresponding Stokes $I$ profiles, normalized to the maximum intensity 
of the Mg~{\sc ii}~$k$ line.
The $V/I$ error bars indicate the $\pm{1}\sigma$ uncertainties resulting from the photon noise.
The WFA fits and the inferred longitudinal magnetic field values are shown in blue  
for the Mn {\sc i} lines, in black for the 
external $V/I$ lobes of Mg~{\sc ii} $h$, and in red for the inner $V/I$ lobes of Mg~{\sc ii} $h$ \& $k$.
\begin{figure}[t]
\begin{center}
\includegraphics[keepaspectratio,height=140mm, angle = 90]{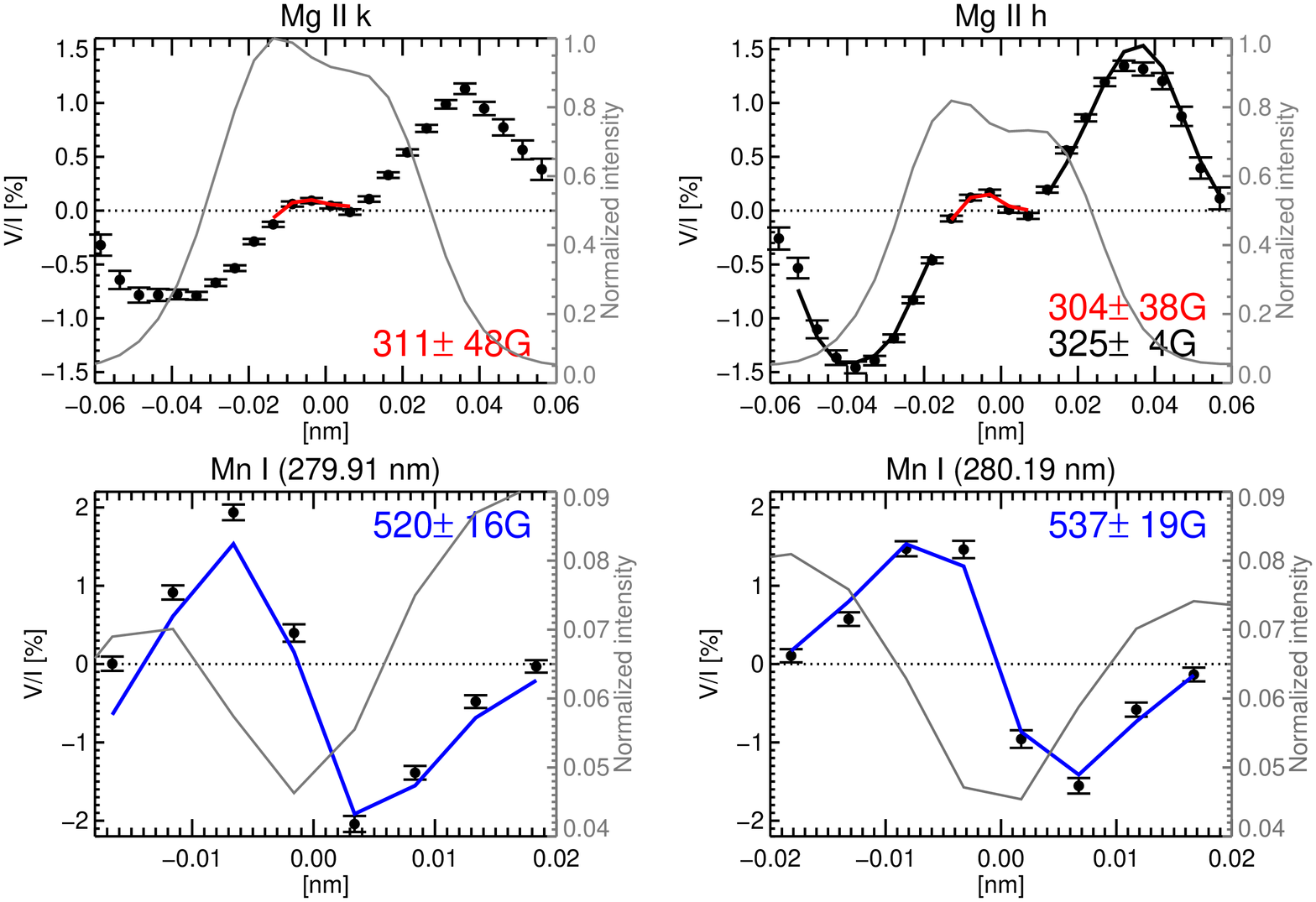}
\end{center}
\end{figure}

\clearpage
\noindent {\bf Fig. 3. Spatial variation of the longitudinal component of the
magnetic field ($B_\mathrm{L}$).}
Blue symbols: $B_\mathrm{L}$ in the lower chromosphere 
determined from the Mn {\sc i} lines. Black symbols: $B_\mathrm{L}$ 
in the middle chromosphere 
determined from the external lobes of the $V/I$ profiles in the Mg {\sc ii} $h$ line. 
Red symbols: $B_\mathrm{L}$ at the top of the upper chromosphere    
determined from the inner lobes of the $V/I$ profiles in the Mg {\sc ii} $h$ \& $k$ lines. 
Green curve: $B_\mathrm{L}$ in the photosphere 
obtained from the Hinode/SOT-SP observations, after spatially smearing the data  
to mimic the CLASP2 resolution (supplementary material 1E).
The error bars of the CLASP2 data represent 1$\sigma$ errors. 
Note that the thin black curves give the normalized
intensity observed by CLASP2
at the $k_{\rm 3}$ line center and at the $k_{\rm 2v}$ 
emission peak of the Mg~{\sc ii} $k$ line 
(solid and dashed-dotted; see the inset of Fig. S4).
The lower panel is a zoom of the upper one.
\begin{figure}[t]
\begin{center}
\includegraphics[keepaspectratio,height=150mm, angle = 90]{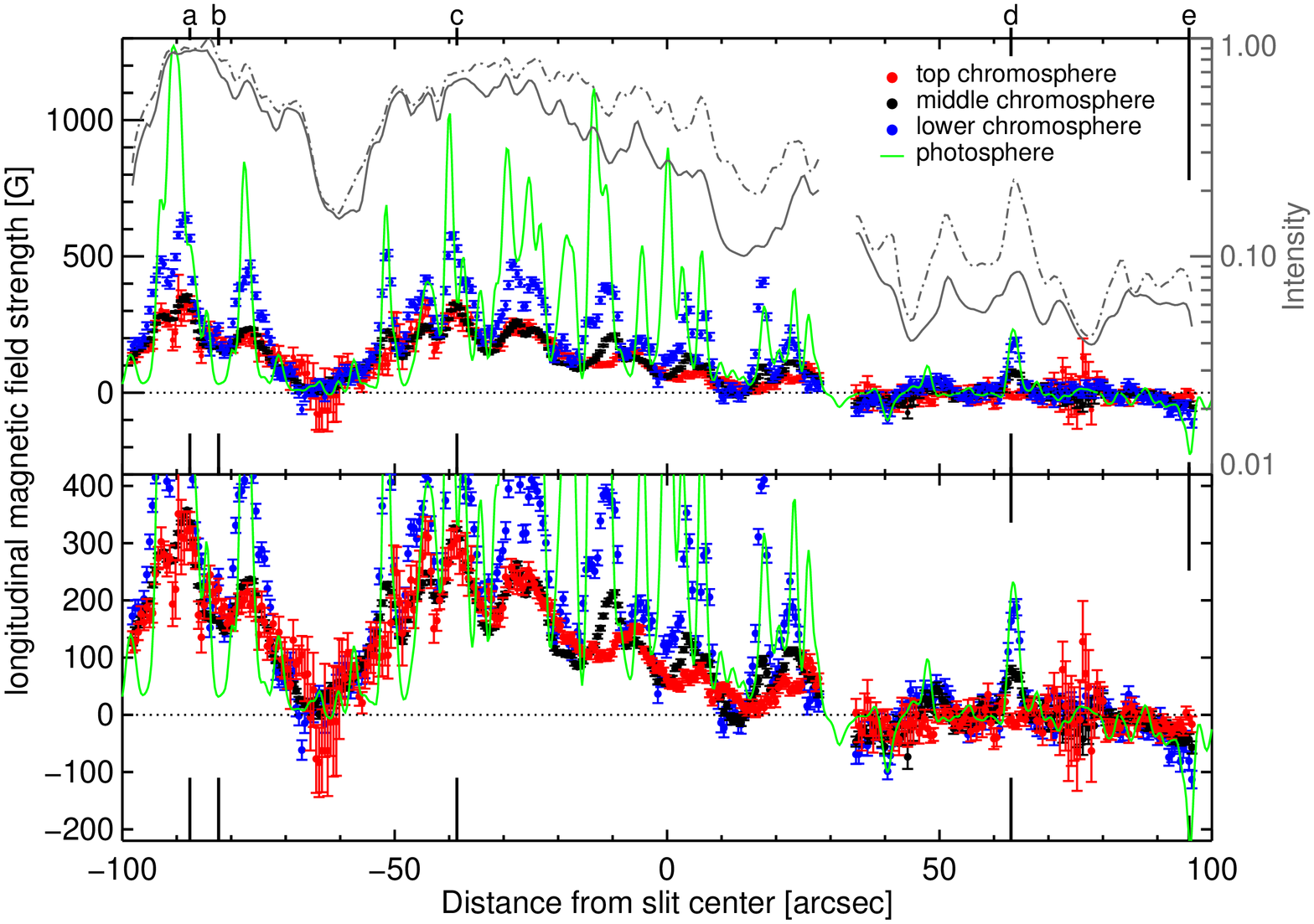}
\end{center}
\end{figure}

\clearpage
\noindent {\bf \rev{Fig. 4.} CLASP2 data corresponding to one of the network features.} 
$V/I$ profiles at location {\bf d} indicated in Fig.~1.
See the caption of Fig.~2 for further explanations.
We find $B_{L}=-5\pm7$~G at the top of the upper 
chromosphere (after averaging the longitudinal field values 
retrieved from the inner $V/I$ lobes of the Mg~{\sc ii} $h$ \& $k$ lines), 
$B_{L}=78\pm7$~G
in the middle 
chromosphere (from the external $V/I$ lobes of Mg~{\sc ii} $h$), 
and $B_{L}=164\pm15$~G 
in the lower chromosphere (from the $V/I$ profiles of the Mn {\sc i} lines).
\begin{figure}[t]
\begin{center}
\includegraphics[keepaspectratio,height=140mm, angle = 90]{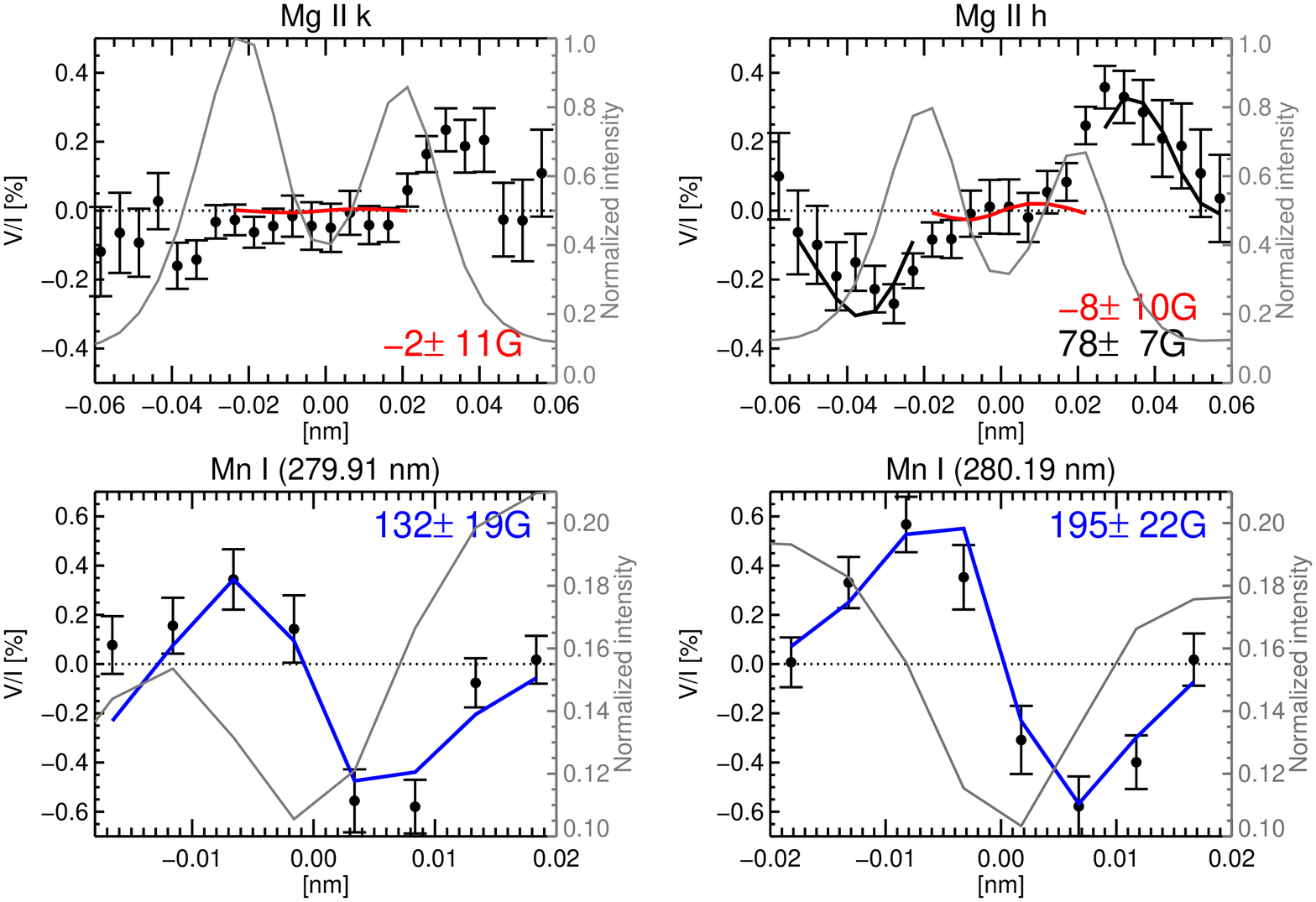}
\end{center}
\end{figure}

\clearpage
\noindent{\bf Table 1. Correlation coefficients (CI for intensity and CP for electron pressure).} 
The table gives the corresponding correlation coefficient 
between the longitudinal component of the 
magnetic field, $B_\mathrm{L}$, inferred at different heights in the atmosphere, and (a) the 
observed Mg~{\sc ii} $k$-line intensity ($k_{\rm 3}$ or $k_{\rm 2v}$)
and (b) the inferred electron pressure near the upper chromosphere (see supplementary material 2E).
The linear Pearson correlation coefficients have been calculated 
considering only the bright plage region pixels located 
between $-98$ and $+28$~arcsec (see Figure 1C).
\begin{table}
\begin{center}
\begin{tabular}{ccccc}
Wavelength window for $B_{\mathrm{L}}$ & Atmospheric layer 
& CI ($k_{\rm 3}$) & CI ($k_{\rm 2v}$) & CP \\ \hline
Inner lobes of Mg~{\sc ii} $h$ \& $k$ & Top of upper chromosphere &0.87 &0.81 &0.74\\
External lobes of Mg~{\sc ii} $h$  & Middle chromosphere & 0.80 &0.75 &0.67\\
Mn~{\sc i} lines around 280~nm & Lower chromosphere & 0.65 &0.63 &0.55\\
Fe~{\sc i} lines around 630.2~nm & Photosphere & 0.48 &0.44 \\ \hline
\end{tabular}
\end{center}
\end{table}

\clearpage

   
\section*{Supplementary Materials}
\rev{Supplementary} Material and Methods\\
Supplementary Text\\
Figs. S1 to S7\\
References \textit{(15-58)}

\clearpage
\section{\rev{Supplementary} Material and Methods}

{\bf 1A. The CLASP2 instrument and the observation}

The Chromospheric Lyman-Alpha Spectro-Polarimeter (CLASP) 
\cite{2013ApOpt..52.8205I,2014SoPh..289.4727I,
2015ApOpt..54.2080N,2015SoPh..290.3081I,2016SoPh..291.3831G,
2017SoPh..292...57G} was designed to measure 
the variation with wavelength of the Stokes $I$, $Q$ and $U$ 
parameters of the hydrogen Lyman-$\alpha$ line at 121.6 nm, where $I$ is the intensity  
\rev{and $Q$ and $U$ the linear polarization.}
This suborbital rocket experiment 
was carried out in September 2015, and provided the first measurement
of the linear polarization signals produced by scattering processes and the Hanle effect 
in the Lyman-$\alpha$ line 
\rev{\cite{kano_discovery_2017,ishikawa_indication_2017,2018ApJ...865...48S,2018ApJ...866L..15T}}.
CLASP2 is the result of a non-negligible modification 
of CLASP, but maintaining its novel design based on 
the inverse Wadsworth configuration \rev{\cite{2020SPIETsuzuki}}. 
The aim was to measure the Stokes $I$, $Q$, $U$ and $V$ 
\rev{(circular polarization)}
parameters  
across the Mg {\sc ii} $h$ \& $k$ lines near 280 nm 
in quiet and active regions of the solar disk with 
a spectral resolution of 0.01~nm, a temporal resolution of 3.2~s  
and a spatial resolution of at least 2 arcsec    
\cite{2018SPIE10699E..30Y,2018SPIE10699E..2WS}. The 
resolutions estimated from the pre-flight measurements
were $\sim$1.1~arcsec and $\sim$0.01~nm. A comparison of the 
spectroscopic data taken by CLASP2 and IRIS during the observation of the plage target   
suggests that these spatial and spectral resolutions
are representative for the CLASP2 flight measurements.

Previous space instruments developed for measuring 
polarization signals in solar ultraviolet lines failed due to in-flight degradation 
problems \cite{stenflo_search_1980},
or lacked the polarimetric sensitivity and resolution 
(spectral, spatial and temporal) needed for accurately determining magnetic fields 
in the upper solar chromosphere and TR of quiet and active regions  
\cite{1981ApJ...244L.127T,
1982SoPh...81..231H,1983SoPh...84...13H,henze_polarimetry_1987,2019ApJ...883L..30M}. 
 
The CLASP2 instrument consists of a Cassegrain telescope with an aperture of 27~cm,
a slit-jaw (SJ) monitor system, and 
a spectropolarimeter to measure the  
Stokes $I$, $Q$, $U$ and $V$ profiles in the wavelength window of the Mg~{\sc ii} $h$ and $k$ lines 
over the 196~arcsec covered by the spectrograph's slit.
The width of the slit corresponds to 0.53~arcsec.
The SJ system is the same one used in CLASP \cite{2016ApJ...832..141K}; 
it takes fast cadence (0.6~s) images of the chromosphere 
using a Lyman-$\alpha$ broad-band filter (FWHM=7~nm). 
The plate scale of the SJ monitor system is 1.03~arcsec per pixel (hereafter, arcsec/pix). 

The spectropolarimeter has two optically symmetric channels, 
which provide simultaneous measurements of two orthogonal polarization states.
The continuously rotating waveplate 
of the Polarization Modulation Unit (PMU) \cite{2015SoPh..290.3081I}
is located in front of the 
slit and modulates the incident radiation (one rotation takes 3.2~s).
At the downstream of the slit, a diffraction grating mounted in 
the inverse Wadsworth configuration serves 
both as a spectral dispersion element and as beam splitter.
Each resulting beam finally goes through a linear polarization analyzer,  
mounted 90$^{\circ}$ from each other, and is 
measured by its corresponding CCD camera.
The two cameras record the modulated intensity every 0.2~s 
in synchronization with the waveplate rotation.
After the dark and gain corrections,
we perform the demodulation and derive the wavelength variation of the four 
Stokes parameters for each channel beam and for each rotation of the waveplate.
Then, we correct for pointing and wavelength drifts and co-register the Stokes spectra. 
Finally we obtain $I$, $Q/I$, $U/I$ and $V/I$ by combining the information from the two channels.
The plate scales of the spectropolarimeter are 0.527~arcsec/pix along 
the slit and 0.00498~nm/pix in the dispersion direction.

CLASP2 conducted the observations at three locations on the solar disk during its ballistic flight.  
From 16:52:53 to 16:53:07 UT on April 11 2019, it observed the center of the solar disk
to verify that the instrumental polarization is negligibly small 
compared with the required polarization accuracy of 0.1\%.
From 16:53:40 to 16:56:16 UT, it observed an active region plage  
which was located at the East side of active region NOAA 12738 (Fig. S1A).
About 130 arcsec of the slit covered part of the bright region of the plage, 
while the remaining 70 arcsec crossed a surrounding 
quieter region including two network elements of a supergranulation cell. 
The data obtained during this 150.4~s had a very stable pointing. 
For the analysis of this paper, in order to increase the signal-to-noise ratio
we have used the temporally averaged spectra from the plage target. 
From 16:56:25 to 16:58:45 UT, CLASP2 observed a quiet region 
near the solar limb, with the slit also radially oriented with respect to the 
center of the solar disk. 

Figure~S2 shows the variation along the slit of the intensity 
(Stokes $I$) and fractional polarization ($Q/I$, $U/I$ and $V/I$) observed 
by CLASP2 in the plage target. The polarization signals theoretically expected for 
the Mg {\sc ii} $h$ \& $k$ lines are clearly seen, 
both the circular polarization produced by the Zeeman effect 
and the linear polarization caused by scattering processes and 
its possible modification by the Hanle and magneto-optical effects
\cite{2012ApJ...750L..11B,2016ApJ...831L..15A,2016ApJ...830L..24D,2020ApJ...891...91D}.
In addition, CLASP2 detected clear circular polarization 
signals produced by the Zeeman effect in Mn {\sc i} lines, as well as in other 
spectral lines which will be the subject of other publications. 
In this paper we focus on the longitudinal component of the magnetic field 
that we can directly infer from the Stokes $I$ and $V/I$ profiles
of the Mg {\sc ii} and Mn {\sc i} resonance lines, while in future investigations 
we will apply a new (presently under development) 
non-LTE inversion technique to the Stokes profiles 
observed by CLASP2 across the Mg {\sc ii} resonance lines.   
 
{\bf 1B. The Hinode/SOT observation}

The Hinode satellite \cite{2007SoPh..243....3K} 
performed coordinated observations of the CLASP2 plage target.
In particular, 
the Spectro-Polarimeter (SP) \cite{2008SoPh..249..233I,2013SoPh..283..579L}
of the Solar Optical Telescope 
(SOT) \cite{2008SoPh..249..167T,2008SoPh..249..197S,2008SoPh..249..221S} 
measured the four Stokes profiles 
of the Fe {\sc i} photospheric lines at 630.15 nm 
and 630.25 nm with a wavelength sampling of 2.16~pm. 

From UT 16:46:57 to 17:51:31
of 2019 April 11 the plage region was scanned using the fast mapping mode, 
with an exposure time of 3.2~s 
at each slit position and a spatial plate scale of 0.32~arcsec/pix. The resulting field of view (FOV) 
was 299~arcsec (solar E-W) by 161~arcsec (solar N-S), which included 
the spatial positions of the slit of the CLASP2 spectrograph (Fig.~1B).
The spatial resolution that can be achieved with the Hinode/SOT-SP is about 0.32~arcsec, with a 
spatial sampling of 0.16~arcsec \cite{2008SoPh..249..167T,2008SoPh..249..197S}, but 
with the chosen fast mapping mode
the spatial resolution was $\sim0.64$~arcsec. 
Thanks to these nearly simultaneous 
coordinated observations with Hinode, we obtained a map of the  
photospheric longitudinal magnetic field strength through 
the application of a Milne-Eddington inversion code
(supplementary material 1E).
 
The co-alignment of the CLASP2 and Hinode/SOT-SP data 
was done using images from 
the Atmospheric Imaging Assembly (AIA) \cite{2012SoPh..275...17L}
and the Helioseismic and Magnetic Imager (HMI) \cite{2012SoPh..275..207S} 
on board the Solar Dynamics Observatory (SDO) \cite{2012SoPh..275....3P}. 
AIA takes full disk images of the solar atmosphere in multiple 
wavelengths with a temporal cadence of 12~s and  
HMI provides full disk measurements of photospheric longitudinal 
magnetic fields (magnetograms) every 45~s,  
both of them with a moderate spatial resolution of $\sim$1.5~arcsec.
The AIA and HMI images are calibrated and co-aligned using 
the data reduction procedure aia\_prep.pro 
in \textit{SolarSoft} (see https://www.lmsal.com/solarsoft/).
The AIA 30.4~nm channel provides images very similar 
to those given by the CLASP2 SJ monitor system \cite{2016ApJ...832..141K}.
Firstly, the co-alignment of the CLASP2 and SDO data was done by 
cross-correlating the AIA 30.4~nm (Fig.~S1A) 
and CLASP2/SJ images (Fig.~S1B).
Secondly, the co-alignment of the SDO and Hinode/SOT images 
was done by cross-correlating  
the photospheric longitudinal magnetic field maps obtained by SOT-SP and HMI. 
The HMI magnetograms were temporally averaged 
over the duration of CLASP2 observations of the plage target.

{\bf 1C. The IRIS observation}

Coordinated observations of the CLASP2 plage target were also carried out with the 
Interface Region Imaging Spectrograph (IRIS) \cite{2014SoPh..289.2733D}, 
from 16:36:45 to 17:51:27 UT of 2019 April 11.
The roll angle of the IRIS spacecraft 
was set to $71^{\circ}$ and the slit of its spectrograph 
was almost parallel to the CLASP2 slit (Fig.~S1).

The IRIS raster scans, each with 16 steps and a 1~arcsec gap 
between adjacent slit positions, cover a FOV
of 15~arcsec (perpendicular to the slit) $\times$ 175~arcsec 
(parallel to the slit), with a cadence of 28~s 
(see the green dashed boxes in Fig.~S1). 
Five raster scans occurred during the CLASP2 observations.
With its spectrograph (SG),  
IRIS measured the Stokes $I$ spectra in two wavelength ranges, 
in the near ultraviolet (NUV) and in the far ultraviolet (FUV), but it  
also provided slit-jaw images (SJIs). 
The co-spatial and co-temporal Mg~{\sc ii} $h$ and $k$ NUV spectra, and 
the SJIs with a broad-band passband of 0.4~nm centered
at 279.6~nm (i.e., around the center of the Mg~{\sc ii} $k$ line, panel C of Fig.~S1)   
were used for comparison with the CLASP2 data.
Binnings along the spatial and dispersion directions were carried out on-board,
yielding spatial and wavelength plate scales of 0.33~arcsec/pix and 5.1~pm/pix for the NUV SG data, 
and a spatial plate scale of 0.33~arcsec/pix for the SJIs, 
which implies a spatial resolution of 0.66~arcsec because of the Nyquist theorem.   
Without binning the IRIS spatial resolution in the NUV 
would be 0.4~arcsec, both for the SG and SJIs \cite{2014SoPh..289.2733D}.

The spatial resolution of the CLASP2 spectra can be mimicked using the IRIS SJI data.
The pseudo SJI at 279.6~nm is made by applying the IRIS passband to the CLASP2 intensity spectra,
while the IRIS SJI is spatially convolved with Gaussians of several widths.
The best cross-correlation between the pseudo SJI and 
the real IRIS SJI was found for a Gaussian FWHM of $\sim$1.2~arcsec (panel D of Fig.~S1).

In order to get a better context of the plage target, with IRIS we also carried out 
dense raster scans before and after
the CLASP2 observation. The scan step size was set to 0.35~arcsec and the 
FOV covered by the SG was 112~arcsec (perpendicular to the slit) 
$\times$ 175~arcsec (parallel to the slit), taking about 30~min to complete one raster scan.
Figure~S3 shows the resulting raster intensity image at the fixed wavelength position 
of the Mg~{\sc ii} $k$ line center, taken 
at the closest observing time to the CLASP2 observation.  
These high spatial resolution IRIS data are helpful to understand the structure of the observed plage
in the upper chromosphere.

\clearpage
{\bf 1D. Derivation of the longitudinal magnetic field component with the WFA}

The weak-field approximation (WFA) is the simplest and 
most direct technique to estimate the longitudinal component of the 
magnetic field from the observed intensity and polarization 
profiles \cite{1989ApJ...343..920J,2004ASSL..307.....L}.  
The WFA is applicable in the weak-field regime, namely 
when the Zeeman splitting $g_\mathrm{eff}\Delta\lambda_{B}$ 
(with $g_\mathrm{eff}$ the effective Land\'e factor) is 
significantly smaller than the line's Doppler width $\Delta\lambda_{D}$. 
These quantities are given by   
\begin{equation}
\Delta\lambda_{B}=\frac{e\lambda^2_{0}}{4\pi m_{e}c}B,
\end{equation}
and
\begin{equation}
\Delta\lambda_{D}=\frac{\lambda_{0}}{c}\sqrt{\frac{2k_{B}T}{m}+\xi^2},
\end{equation}
where $e$ is the absolute value of the electron charge, 
$m_{e}$ the mass of the electron, $\lambda_{0}$ the wavelength of the spectral line, 
$c$ the speed of light, $T$ the plasma temperature, $k_{B}$ the Boltzmann's constant, 
$m$ the mass of the atom, and $\xi$ the non-thermal velocity.
Therefore, 

\begin{equation}
\rev{{\cal R}}\,=\,g_\mathrm{eff}\frac{\Delta\lambda_{B}}{\Delta\lambda_{D}}
\propto g_\mathrm{eff}\frac{\lambda_{0}B}{\sqrt{2k_{B}T/m+\xi^2}}.
\end{equation}

In the chromosphere, where the UV spectral lines here considered originate,  
the temperature and the non-thermal velocity are larger and the magnetic field 
strength is weaker than in the underlying photosphere. 
\rev{Typically, ${\cal R}{\approx}0.06$ for the Mg~{\sc ii} lines
(assuming $T=8000$~K and $B=300$~G) and ${\cal R}{\approx}0.3$ 
for the Mn~{\sc i} lines (assuming $T=4000$~K and $B=500$~G) 
when the non-thermal velocity $\xi$ is not  
included in the estimation (see supplementary materials 2C and 2D for the
effective Land\'e factors). Given that in the chromosphere $\xi$ is very significant, in reality
such $\cal R$ values are even smaller (i.e., the weak field regime holds for both spectral lines).}

In the weak-field regime, the following equation is 
strictly valid when the line-of-sight component of the magnetic field 
is constant in the atmospheric region where the Stokes $V$ 
profile of the spectral line under consideration is formed:
\begin{equation}
V(\lambda) = - \frac{e\lambda^2_{0}}{4\pi m_{e}c}g_\mathrm{eff}B_\mathrm{L}\biggl(\frac{\partial I}{\partial \lambda}\biggr)=-4.67\times10^{-12}g_{\mathrm{eff}}\lambda^{2}_{0}B_{\mathrm{L}}\biggl(\frac{\partial I}{\partial \lambda}\biggr), \label{eq:wfa}
\end{equation}
where $\lambda_{0}$ is in nm and $B_\mathrm{L}=B{\rm cos}{\theta_B}$ is the longitudinal 
component of the magnetic field,
with $B$ the magnetic strength in gauss and $\theta_{B}$ the angle between 
the magnetic field vector and the line of sight (LOS). 
After calculating the derivative of the Stokes $I$ profile with respect to wavelength, the 
scaling factor that provides the best fit to the circular polarization profile is obtained  
by means of the least-square minimization method.
The 1-$\sigma$ uncertainties of the retrieved $B_\mathrm{L}$ values 
were estimated from the $V(\lambda)$ measurement errors, the main source of which 
is the photon noise.

As is well-known, the magnetic field in the solar atmosphere is not 
constant along the LOS. Nevertheless, as discussed in supplementary material 2C,
the WFA is still applicable to estimate the longitudinal field component 
in the line formation region,
provided that the magnetic field is in the weak field regime and that the spatial 
gradients of the LOS component of the magnetic 
field are not too extreme in the region of formation of the Stokes $I$ and $V$ 
spectral features.

As mentioned in supplementary material
1A, in order to increase the signal-to-noise ratio we have averaged 
the Stokes profiles over the 150.4 s duration of the plage-target observation. Although the 
solar chromosphere is very dynamic, with significant variations of the macroscopic velocity and of 
the kinetic temperature and density of the plasma over shorter time scales, its magnetic structure 
is expected to be much simpler than its thermodynamics, especially 
in the $\beta{<}1$ regions where it is almost force-free, and to show a relatively slow time 
evolution \cite{2006ASPC..354..259J}. 
The important question is whether the longitudinal field component inferred from the 
temporally-averaged Stokes profiles is significantly affected by their temporal evolution, 
mainly dominated by the dynamics and thermodynamics of the chromospheric plasma.
We have checked that the wavelength derivative 
of the observed intensity profiles of the Mg~{\sc ii} $h$ \& $k$ lines at each temporal  
step of 3.2~s shows similar patterns and does
not lead to any significant cancellation 
or reduction of the temporally-averaged signals.

We point out that what the WFA provides is  
the average longitudinal field strength over the resolution 
element (i.e., the net magnetic flux divided by the area of the resolution element).
This quantity is equal to the longitudinal magnetic field strength if the magnetic filling factor ($f$)   
is unity in the atmospheric region of formation of the considered spectral line interval  
(in a two-component model $f$ is 
the fraction of the resolution element covered by the magnetic component).
Because of the rapid expansion of the magnetic field in active region plages  
the magnetic filling factor rapidly approaches unity as one goes to higher atmospheric layers. 
We assume $f=1$ in our inferences from the CLASP2 data. 

{\bf 1E. Determination of the photospheric magnetic field from the Hinode/SOT-SP data}

The photospheric magnetic field in plage regions is generally strong, of the order of kG, 
and thus the WFA is not generally applicable to infer the magnetic field from the Fe {\sc i} lines 
around 630~nm. Therefore, in order to obtain the LOS component of the magnetic field \rev{in the 
photosphere of the observed plage region, } 
we have used the SOT-SP Level 2 information given by  
the Community Spectro-polarimetric Analysis Center 
(CSAC) of the High Altitude Observatory (DOI:10.5065/D6JH3J8D. See also \href{https://www2.hao.ucar.edu/csac}{https://www2.hao.ucar.edu/csac}).
This information stems from the application of the MERLIN Milne-Eddington
(ME) inversion code to the SOT-SP Level 1 data (DOI:10.5065/D6T151QF), 
which provides the magnetic field vector 
(field strength $B$, inclination $\theta_B$, and azimuth $\chi_B$), 
the line-of-sight velocity, the source function, the Doppler broadening as well as the macroturbulence 
and a stray light factor. To this end, MERLIN assumes a single atmospheric 
magnetic component and fits the observed intensity profile   
with $fI_{mag}+(1-f)I_{stray}$, where $I_{mag}$ is the intensity from the magnetic component and 
$I_{stray}$ the stray light profile. The fraction $(1-f)$ of stray light 
needed to fit the observed intensity spectrum varies between 0 and 1, and the 
filling fraction $f$ can in principle be interpreted as the fraction of the pixel 
occupied by the assumed single atmospheric magnetic component. 
The $f$ values provided by MERLIN at the spatial points 
of the plage crossed by the CLASP2 slit 
vary between 0.2 and 0.95
(see the brightest region in Fig. 1A). The quantity 
$B\cos{\theta_B}$ is a good estimation of the longitudinal field strength $B_\mathrm{L}$ 
if the assumption of a single atmospheric magnetic component within each pixel 
is reasonable and if there is no significant degeneracy between the values inferred 
for $f$ and $B$. In this paper, we have assumed this to be the case, 
reason why Figure ~1B and the green curve of Figure 3 show $B\cos{\theta_B}$. 

The Hinode/SOT-SP data has better spatial resolution 
than that of the CLASP2 data.
In order to properly compare the spatial variation of the 
photospheric longitudinal magnetic field, 
$B_\mathrm{L}$, with the ones we have obtained from the 
Stokes $I$ and $V/I$ profiles observed by CLASP2, 
we have smeared the $B_\mathrm{L}$ map obtained from the SOT-SP observations to 
mimic the CLASP2 spatial resolution.
The convolution of the IRIS data with a Gaussian with a FWHM of 1.2~arcsec provides 
the best cross-correlation with the CLASP2 data (supplementary material 1C). Because the spatial 
resolution of the SOT-SP and IRIS instruments are comparable, we have applied 
the same gaussian smoothing to the SOT-SP map (see the green curve of Fig. 3).

\section{Supplementary Text}

{\bf 2A. Intensity spectra}

Figure~S4 shows the temporally and spatially 
averaged intensity spectra observed by CLASP2 outside the bright part of the plage target, 
which covers 60~arcsec (see Fig. 1C).  \rev{The investigation of 
this paper is based on the Stokes $I(\lambda)$ 
and $V(\lambda)/I(\lambda)$ profiles of the Mn {\sc i}   
and Mg~{\sc ii} lines indicated in Figure S4 (hereafter, Stokes $I$ and $V/I$). }

The Stokes $I$ profiles of the Mg~{\sc ii} $h$ and $k$ lines show the well-known 
two peak structure (see the inset of Fig. S4), 
where we distinguish three features:
$k_{3}$ (line center depression), $k_{2\mathrm{v}}$ and $k_{2\mathrm{r}}$ 
(the violet and red emission peaks),
and $k_{1\mathrm{v}}$ and $k_{1\mathrm{r}}$ (the violet and red minima) 
of the Mg~{\sc ii} $k$ line (for the 
Mg~{\sc ii} $h$ line the same features are dubbed $h_{3}$, 
$h_{2\mathrm{v}}$, $h_{2\mathrm{r}}$, $h_{1\mathrm{v}}$, and $h_{1\mathrm{r}}$).

In Figure~3 and Table~1, we provide information at the 
$k_{3}$ and $k_{2\mathrm{v}}$ wavelengths. 
At most of the pixels along the slit, 
the $k_{2\mathrm{v}}$ intensity peaks were successfully identified. 
However, in some pixels of the bright region of the plage target, 
the Mg~{\sc ii} $h$ \& $k$ intensity spectra showed blue-shifted single-peak profiles, without  
any clear $k_{3}$ or $h_{3}$ line-center depression. 
For such pixels we have used the $k_{3}$ wavelength of the spatially averaged intensity profile.

{\bf 2B. Detailed description of the circular polarization observed in Mg~{\sc ii} $h$ and $k$}

Figure S5 shows the intensity and circular polarization profiles of the Mg~{\sc ii} and 
Mn~{\sc i} lines at the location {\bf b} indicated in Figure~1. 
The Mg~{\sc ii} $h$ line shows larger $V/I$ amplitudes 
than the Mg~{\sc ii} $k$ line ($g_\mathrm{eff}=4/3$ and $7/6$, respectively), 
as expected from their effective Land\'e factors.  
The $V/I$ profiles of the Mg~{\sc ii} resonance lines show four lobes, both 
inside and outside the bright region of the observed plage,  with the inner lobes (located 
just around the line centers, $\Delta{\lambda}<0.015$~nm) 
presenting significantly smaller amplitudes than the external ones. 
A comparison between the intensity (gray line in Fig.~S5) 
and the $V/I$ profiles indicates that
the inner lobes extend from the $k_{3}$ ($h_{3}$) center till 
the $k_{2\mathrm{v}}$ and $k_{2\mathrm{r}}$ ($h_{2\mathrm{v}}$ 
and $h_{2\mathrm{r}}$) peaks of the intensity profiles, 
while the external lobes appear in the outer wings of 
the $k_{2\mathrm{v}}$ and $k_{2\mathrm{r}}$ ($h_{2\mathrm{v}}$ and $h_{2\mathrm{r}}$) peaks. 
\rev{Figure~S5 also shows the best fits produced by the wavelength 
derivative of the intensity profile, which is also normalized 
to $I(\lambda)$ (i.e., to the intensity at each spectral pixel).}
We point out that the relatively small amplitudes of the inner lobes 
of the plage $V/I$ profiles are caused by the corresponding flattened intensity profiles in the 
spectral region between $k_{2\mathrm{v}}$ ($h_{2\mathrm{v}}$) 
and $k_{2\mathrm{r}}$ ($h_{2\mathrm{r}}$) peaks.

{\bf 2C. Application of the WFA to the Mg~{\sc ii} $h$ and $k$ lines}

Radiative transfer calculations in semi-empirical models of the solar atmosphere
with constant magnetic fields weaker than 1000 G 
show that, while the WFA provides an excellent fit to the inner 
lobes of the $V/I$ profiles of the Mg~{\sc ii}~$k$ line 
($g_\mathrm{eff}=7/6$), it overestimates 
the amplitudes of the external lobes due to the impact of 
partial frequency redistribution (PRD) on the fractional circular polarization   
\cite{2016ApJ...830L..24D,2016ApJ...831L..15A}.
This implies that if the conditions under which the WFA hold are satisfied, 
then its application to the inner lobes alone provides a reliable 
estimation of $B_\mathrm{L}$ in the region of the solar 
atmosphere where these lobes originate, while its application only to the 
external lobes would underestimate $B_\mathrm{L}$ in the corresponding 
atmospheric region of formation. The same is valid for the 
Mg~{\sc ii}~$h$ line ($g_\mathrm{eff}=4/3$), but the underestimation of the $B_\mathrm{L}$ values 
that result from the application of the WFA to the external 
lobes of this line is less significant (i.e., the use of the $V/I$ 
external lobes of the $h$ line leads to an error of about 10\%, while it is a bit   
larger for the $k$ line). 
 
There is also a blend in the external blue wing of the $k$ line.  
Thus, in this paper we have applied the WFA separately to the inner lobes of 
the $V/I$ profiles of $h$ and $k$ and 
to the external lobes of the $V/I$ profiles of only the $h$ line.

In the solar atmosphere the magnetic field changes along the LOS, 
but if this variation is not too strong the WFA still provides reliable estimations of the 
longitudinal field component. We illustrate this in Figure S6, the right panel of which 
shows the results of PRD radiative transfer calculations  
in a semi-empirical model of the quiet solar  
atmosphere \cite{1993ApJ...406..319F} 
permeated by a vertical magnetic field with a strength varying 
exponentially with height as 
indicated in the left panel. The black filled circles show the calculated $V/I$ profile, while the 
solid curves show the fits obtained when applying the WFA only to the external $V/I$ lobes 
(gray curves) and only to the inner $V/I$ lobes (green curve). 
The longitudinal magnetic field 
strength inferred from the inner lobes is $B_{\rm c}=15$~ G (this value corresponds to 
the top of the model's upper chromosphere; see the green circle in the left panel). The application of 
the WFA to the external lobes gives instead $B_{\rm w}=158$~G, which corresponds to the middle 
chromosphere (see the gray circle in the left panel).
These spatial locations are in agreement with the information provided by the contribution function 
of the inner and external lobes of the $V/I$ profile, as shown by the colored 
regions of Figure~S6. The same conclusion, namely that the inner lobes of the $V/I$ profiles of 
the Mg {\sc ii} $h$ \& $k$ lines provide information on the magnetic field at the  
top of the upper chromosphere and the external lobes at 
middle chromospheric heights, is confirmed by the response function of $V/I$ to magnetic field 
perturbations in semi-empirical models of the solar atmosphere \cite{2020ApJ...891...91D}. 
The just quoted paper, which explains how the calculations of Figure S6 were done,  
shows also that in a plage model 
the response function of the external lobes peaks closer to the upper chromosphere than 
in a quiet Sun model.

On average, the Mg~{\sc ii} $k$ line forms only about 50 km 
higher than the Mg~{\sc ii} $h$ line \cite{2012ApJ...750L..11B,2013ApJ...772...90L}, 
a distance significantly smaller than the pressure scale height in the solar chromosphere. 
The red symbols in Figure~3 show the $B_\mathrm{L}$
values that result from averaging those obtained independently from the inner $V/I$ lobes 
of the Mg {\sc ii} $h$ \& $k$ lines. From the information provided above,  
the magnetic field strengths given by the red symbols in 
Figure 3 can be associated to the top of the upper chromosphere (i.e., near the TR), 
while the black symbols to the middle chromosphere.

{\bf 2D. Application of the WFA to the Mn~{\sc i} lines}

The CLASP2 spectral window includes three resonance lines 
of Mn {\sc i} resulting from transitions 
between the ${\rm a}^6{\rm S}_{5/2}$ level of the ground term 
and the following upper levels of increasing energy: 
${\rm y}^6{\rm P}_{3/2}$, ${\rm y}^6{\rm P}_{5/2}$ and ${\rm y}^6{\rm P}_{7/2}$. 
One of these lines lies in the blue near 
wing of Mg {\sc ii} $k$, very close to the $k_{1\mathrm{v}}$ 
wavelength, while the other two lines lie in the far wing between 
the Mg {\sc ii} $k$ \& $h$ line cores (see Fig. S4). 
These two lines at 279.91~nm and 280.19~nm are the Mn {\sc i} lines we have 
used in this paper, because they show very clear 
circular polarization signals produced by the Zeeman effect (see Fig. 1), 
each line showing two $V/I$ lobes.  

A non-LTE radiative transfer investigation  
that will be published elsewhere shows that the $V/I$ signals of such Mn {\sc i} lines  
originate in the lower chromosphere, near the 
temperature minimum region in standard solar semi-empirical models. 
Accordingly, the lower chromosphere is the approximate region of formation 
we assign to the blue-symbol values in Figure~3. 
Manganese has nuclear spin $I=5/2$ and the same theoretical  
investigation reveals that the ensuing hyperfine 
structure (HFS) with the crossings and repulsions between the hyperfine F-levels 
have a significant impact on the Stokes $I$ and $V$ 
profiles of the Mn {\sc i} lines observed by CLASP2. 
The inclusion of HFS widens the Stokes $I$ profiles and lowers 
the amplitude of Stokes $V/I$ profiles while also separating their peaks. Moreover, 
that investigation shows that the WFA is applicable to the two Mn {\sc i} lines we have selected, 
when using $g_{\mathrm{eff}}=1.94$ for the 279.91~nm line 
and $g_{\mathrm{eff}}=1.7$ for the 280.19~nm 
line (these are the fine-structure effective Land\'e factors). 

{\bf 2E. Thermodynamical parameters inferred from CLASP2 intensity spectra}

As shown in Table~1, there appears to be a good correlation between the magnetic field 
measured with CLASP2 and the Mg~{\sc ii} $k_\mathrm{3}$  and $k_\mathrm{2v}$ 
intensities and, as pointed out 
in the main text, the intensity in the $k$ line-core spectral features
correlates with the temperature 
at its formation height \cite{2013ApJ...772...90L}. 
To further investigate the relationship between the magnetic field and the local thermodynamics properties of the chromosphere, we exploit the IRIS$^2$ database \cite{2019ApJ...875L..18S}. The latest version of this database contains 150,000 spectral profiles around the Mg~{\sc ii} $h$ \& $k$ lines, and the associated model atmospheres (i.e., temperature, velocity, turbulence, electron density, as a function of optical depth), calculated using the STiC inversion code which includes the effects of radiative transfer with PRD \cite{2019A&A...623A..74D}. 

For each CLASP2 Stokes $I$ spectral line profile of the Mg~{\sc ii} $h$ \& $k$ lines, we determine the profile in the IRIS$^2$ database that has the smallest weighted Euclidean distance to the CLASP2 profile. Since we are interested mostly in the chromospheric properties, we place enhanced weighting on the wavelength regions between Mg~{\sc ii} $k_\mathrm{1v}$ and Mg~{\sc} $k_\mathrm{1r}$
and between Mg~{\sc ii} $h_\mathrm{1v}$ and Mg~{\sc} $h_\mathrm{1r}$, as well as the Mg~{\sc ii} triplet line formed around 279.89~nm, which are all formed in the chromosphere. The associated model atmosphere from IRIS$^2$ is then utilized to investigate its correlation with the CLASP2 magnetic field measurements. 
We find that in the middle and in the upper chromosphere 
the electron pressure (i.e., the product of temperature 
and electron density) shows a high cross-correlation with the magnetic field (see Table 1).   
 Figure~S7 shows the product of temperature and electron density along the CLASP2 slit, as derived from the IRIS$^2$ inversions, in comparison with the CLASP2 magnetic field measurements. 
The cross correlation values are similar to those of the Mg~{\sc ii} $k_\mathrm{3}$ 
and $k_\mathrm{2v}$ intensities shown in Figure 3.
The high correlation with the electron pressure provides support for a close relationship between magnetic field strength in the chromosphere and local heating. This is because heating leads to an increase in local temperature, density, and ionization.

We note that the IRIS$^2$ database was created for spectral line profiles at the spatial resolution (0.4 arcsec) and temporal resolution ($2-30$~s) of IRIS. The CLASP2 intensity measurements are averaged over 150.4~s with a lower spatial resolution. Such averaging generally leads to smoother spectral line profiles since the spatio-temporal averaging tends to smear out subtle spectral features. To find the best matches with the CLASP2 observations we thus first perform spectral smoothing on the IRIS$^2$ profiles so that the profiles show a similar level of detail as those of the CLASP2 measurements.

Inversions of complex chromospheric spectral lines formed under 
non-LTE conditions have become possible thanks to advances in numerical radiative transfer  
 and inversion techniques \cite{2000ApJ...530..977S}.
The derived thermodynamic parameters are approximations and not necessarily unique solutions to the underlying atmospheric conditions. 
In addition, the inherent properties of the radiative transfer in spectral lines 
imply that they are sensitive to local conditions at varying levels, depending on the height range considered 
\cite{2018A&A...620A.124D,2020A&A...634A..56D}.
Despite these inherent limitations (which are not always easy to quantify), tests in which inversion results are compared with ``ground truth'' from numerical simulations suggest that they can provide valuable information about the atmosphere responsible for the observed emissions  \cite{2012A&A...543A..34D,2019A&A...623A..74D}. 

With these caveats in mind, our results suggest that there exists a correlation between two independent ``measurements'' in the atmosphere: the line-of-sight magnetic field (as derived from the Stokes $V$ CLASP2 measurements), and the product of temperature and electron density (as derived from comparison with STiC inversions in the IRIS$^2$ database).

\clearpage
\begin{figure}[h]
\begin{center}
\includegraphics[keepaspectratio,height=130mm]{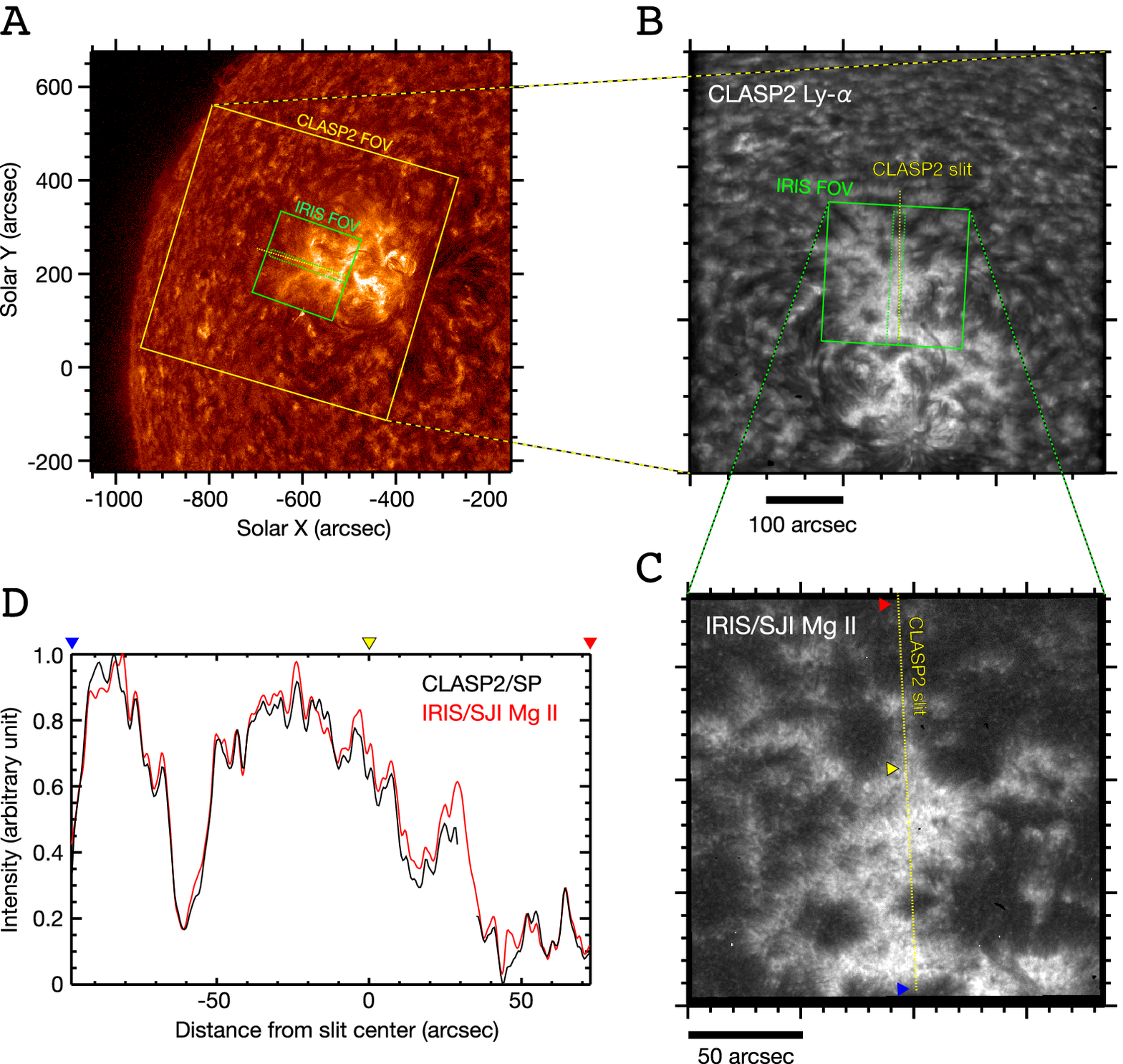}
\end{center}
\end{figure}
\noindent {\bf Fig. S1. CLASP2 plage target.}  
{\bf (A)} SDO/AIA 30.4 nm image taken at 16:54:29 UT 
on 2019 April 11.
{\bf (B)} Simultaneous CLASP2/SJ image.
{\bf (C)} Simultaneous IRIS SJI at 279.6~nm.
{\bf (D)} Comparison between the pseudo SJI made from the CLASP2 spectra (black)
and the IRIS SJI at 279.6~nm (red).
The IRIS SJI is smoothed with a Gaussian of FWHM = 1.2~arcsec, 
which provides the highest correlation with the pseudo SJI.
The yellow solid box represents the CLASP2/SJ FOV of the 
 the plage target, while the yellow solid line indicates the CLASP2 slit.
The green dashed and solid boxes represent the FOVs observed by IRIS SJ and SG.

\clearpage
\noindent {\bf \rev{Fig. S2.} Stokes spectra observed by CLASP2 in the plage target.} 
Each Stokes parameter was temporally averaged over the 150.4~s observing time.
We point out that while the $V/I$ signals are due to the Zeeman effect in 
several lines contained in the CLASP2 spectral window,   
the $Q/I$ and $U/I$ signals are dominated by the scattering of anisotropic radiation 
and the Hanle and magneto-optical effects across the Mg {\sc ii} $h$ \& $k$ lines
\cite{2012ApJ...750L..11B,2016ApJ...830L..24D,2016ApJ...831L..15A,2020ApJ...891...91D}.   
\begin{figure}[t]
\begin{center}
\includegraphics[keepaspectratio,width=120mm]{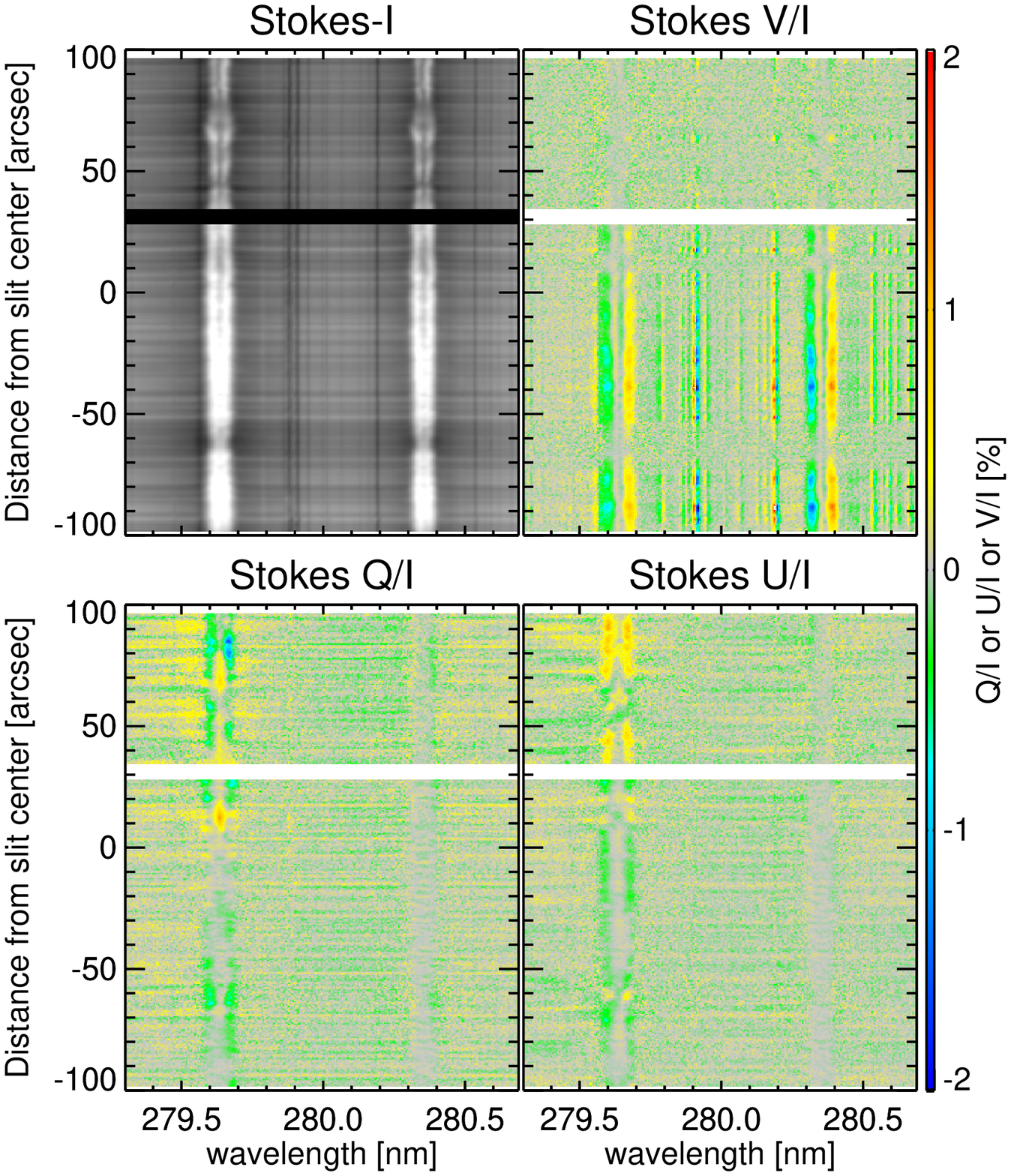}
\end{center}
\end{figure}

\clearpage
\begin{figure}[t]
\begin{center}
\includegraphics[width=90mm]{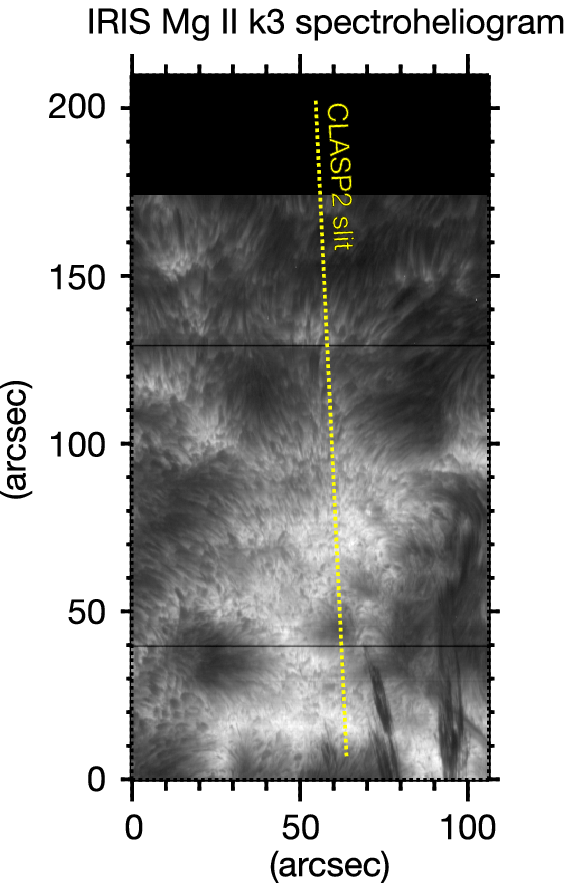}
\end{center}
\end{figure}
\noindent {\bf Fig. S3. Synoptic view of a dense IRIS raster scan of the plage target.}  
Intensity image at the fixed wavelength position of the Mg{\sc ii} $k$ line center,  
taken on 2019 April 11 from 15:57:54 to 16:25:26 UT.
The yellow line shows the approximate position of the CLASP2 slit.

\clearpage
\noindent {\bf Fig. S4. Temporally and spatially averaged 
intensity observed by CLASP2 outside the plage.} 
The spectral lines used in this paper are indicated.
The inset shows the well-known features of the Mg~{\sc ii} $k$ line.
The vertical axis shows the averaged number of electrons per 3.2~s (one PMU rotation).
\begin{figure}[t]
\begin{center}
\includegraphics[keepaspectratio,height=160mm, angle = 90]{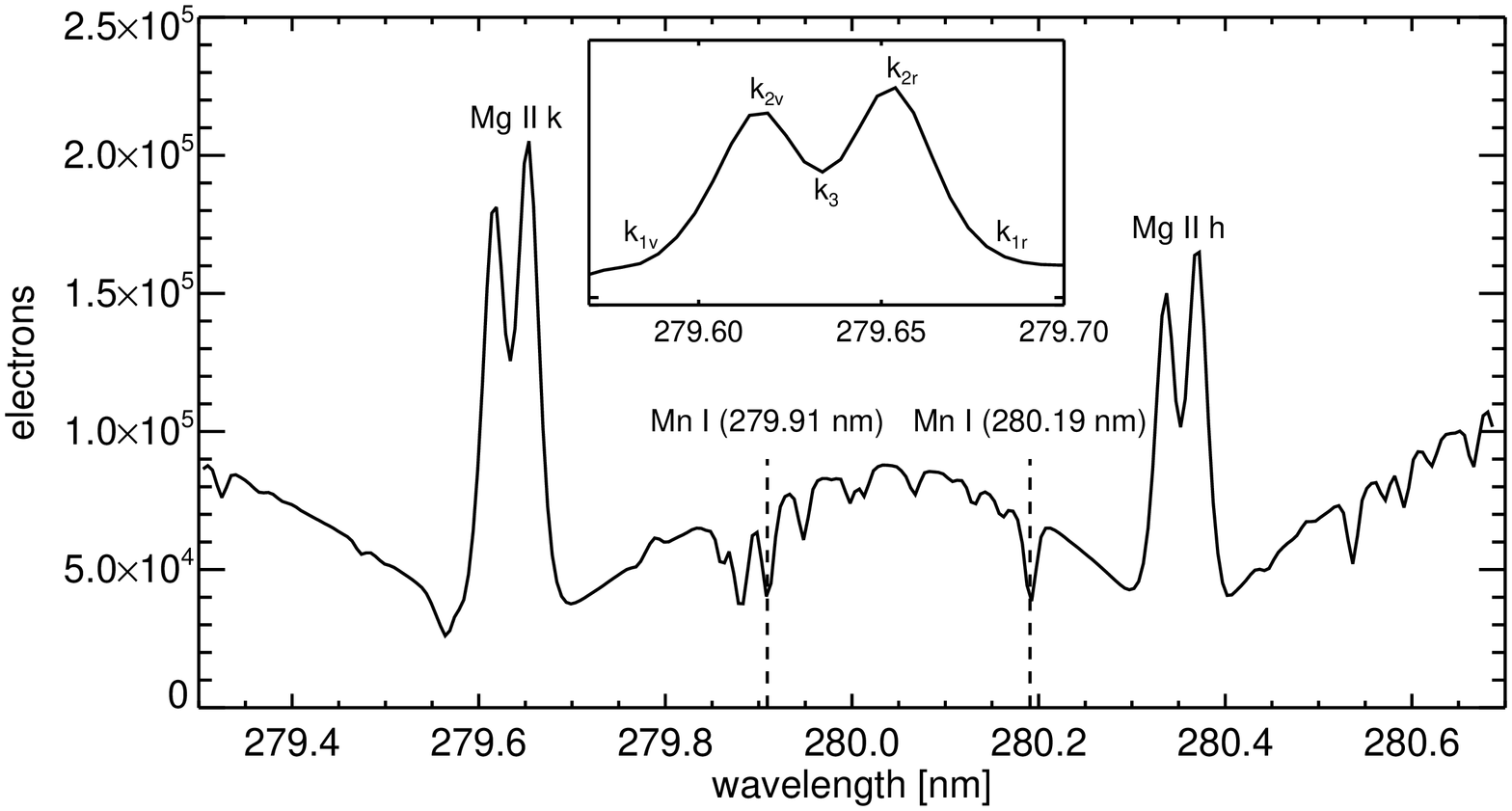}
\end{center}
\end{figure}

\clearpage
\noindent {\bf \rev{Fig. S5.} The $I$ and $V/I$ profiles observed by CLASP2 at location {\bf b} of Fig. 1.} 
$V/I$ profiles of Mg {\sc ii} $k$ at 279.64~nm, Mg {\sc ii} $h$ at 280.35~nm, 
Mn {\sc i} at 279.91~nm, and Mn~{\sc i} at 280.19~nm
at the location {\bf b} indicated in Figure~1, where the $B_\mathrm{L}$ 
values in the photosphere are smaller than in the chromosphere. 
\rev{The circular polarization $V(\lambda)$ is normalized to $I(\lambda)$ 
(i.e., to the intensity at each spectral pixel).} 
The gray curves show the
corresponding Stokes $I$ profiles, normalized to the maximum intensity of Mg~{\sc ii}~$k$.
The $V/I$ error bars indicate the $\pm{1}\sigma$ uncertainties resulting from the photon noise.
The WFA fits and the inferred longitudinal magnetic field values are shown in blue  
for the Mn {\sc i} lines, in black for the 
external $V/I$ lobes of Mg~{\sc ii} $h$, and in red for the inner $V/I$ lobes of Mg~{\sc ii} $h$ \& $k$.
\begin{figure}[t]
\begin{center}
\includegraphics[keepaspectratio,height=160mm, angle = 90]{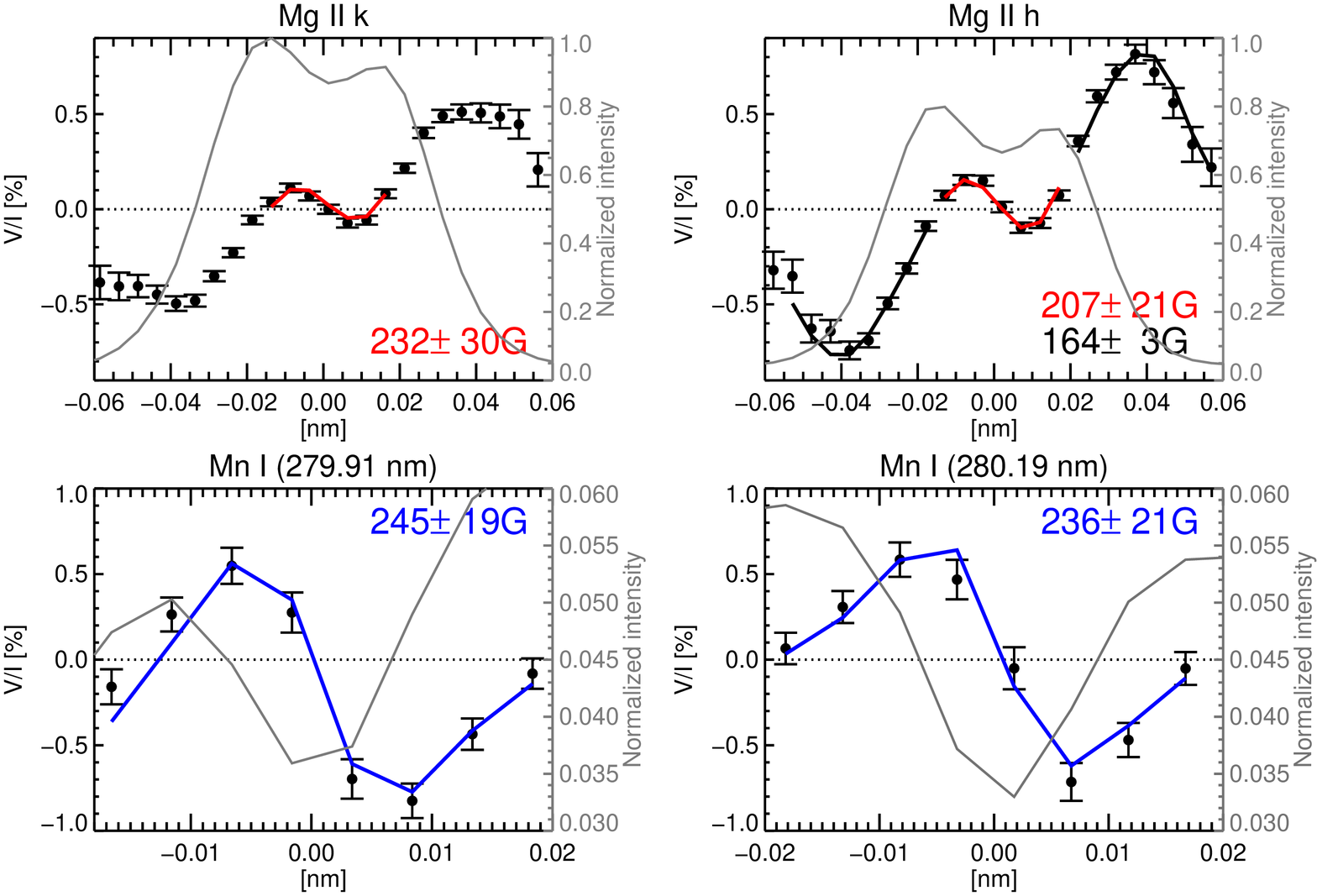}
\end{center}
\end{figure}

\clearpage
\begin{figure}[h]
\begin{minipage}{0.5\hsize}
  \begin{center}
   \includegraphics[width=85mm]{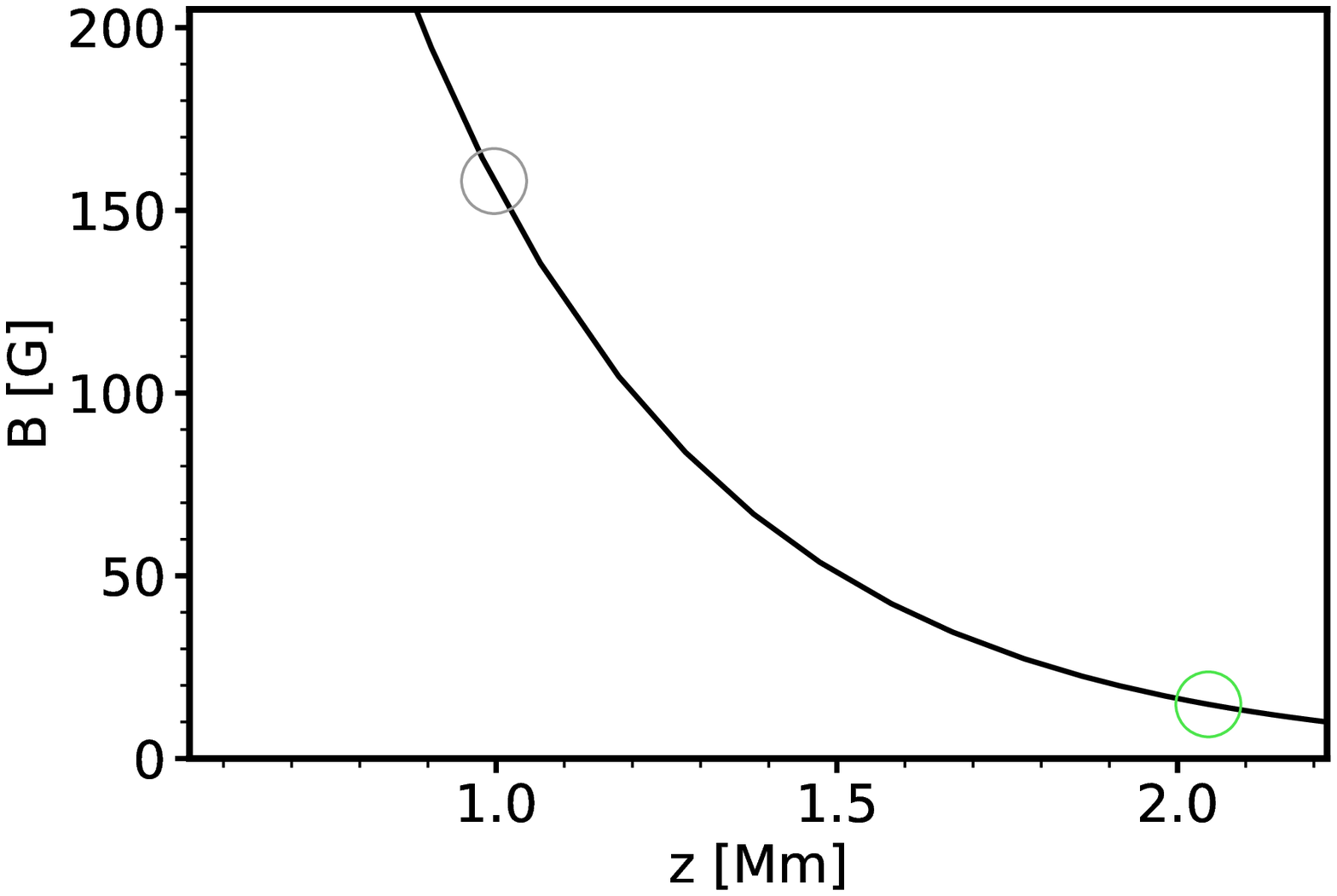}
  \end{center}
 \end{minipage}
 \begin{minipage}{0.5\hsize}
  \begin{center}
   \includegraphics[width=85mm]{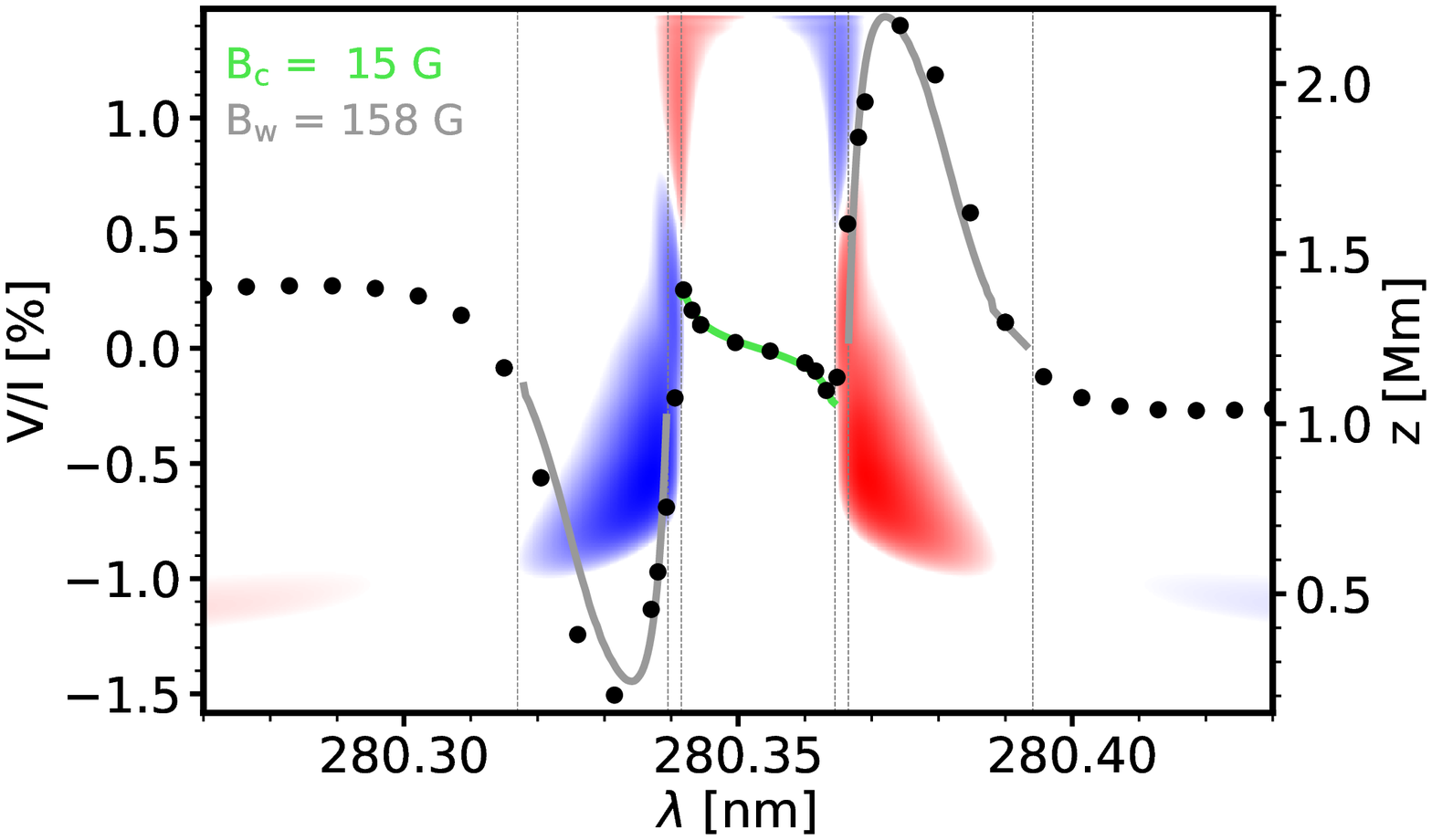}
  \end{center}
 \end{minipage}
\end{figure}
\noindent {\\\\\\ \bf Fig. S6. Performance of the WFA with height-dependent magnetic fields.} 
Left panel: variation with height of the strength of a vertical magnetic 
field in a semi-empirical model of the quiet solar atmosphere \cite{1993ApJ...406..319F}.  
Right panel: the black filled circles show the $V/I$ profile of the Mg {\sc ii} $h$ line 
calculated at the model's disk center, while the 
solid curves show the fits obtained when applying the WFA only to the external $V/I$ lobes 
(gray curves, which give $B_{\rm w}=158$~G) and only to the inner $V/I$ lobes (green curve, which gives 
$B_{\rm c}=15$~G). The colored areas show the $V/I$ contribution function, which indicate 
that the inner lobes originate at the top of the model's upper 
chromosphere (green open circle in the left panel) 
and the external lobes in the middle chromosphere (gray open circle in the left panel).

\clearpage
\noindent {\bf Fig. S7. Spatial variation of the longitudinal magnetic field 
and of the product of temperature and electron density.} 
The purple curve shows the product of 
temperature and electron density (i.e., the electron pressure) near the 
upper chromosphere (i.e., at $\log\tau_{500}=-5.4$,  
with $\tau_{500}$ the continuum optical depth at 500 nm).   
The blue and red symbols 
show the LOS component of the magnetic field in the lower chromosphere 
and at the top of the upper chromosphere, respectively, as in Fig.~3.
\begin{figure}[t]
\begin{center}
\includegraphics[keepaspectratio,height=160mm,angle = 90]{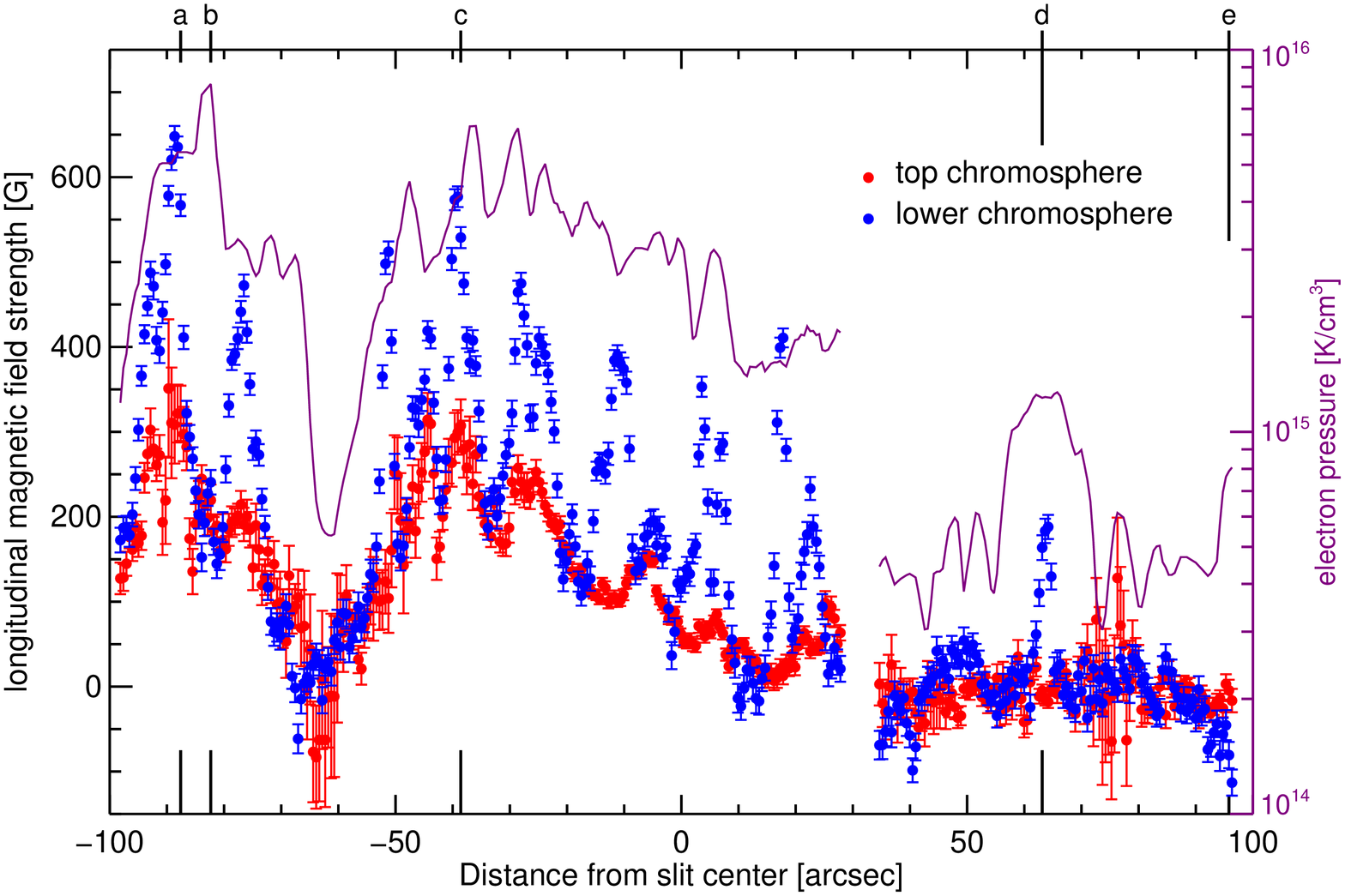}
\end{center}
\end{figure}

\end{document}